\documentclass[aip,jcp,superscriptaddress,amsfonts,graphicx,preprint]{revtex4-2}
\usepackage[utf8]{inputenc}
\usepackage[T1]{fontenc}
\usepackage{tgtermes}
\usepackage{amsmath, amssymb}
\usepackage{mathtools}
\usepackage{breqn}
\usepackage{graphicx}
\usepackage{float}
\usepackage{bm}
\usepackage{physics}
\usepackage{xcolor}
\usepackage{verbatim}
\usepackage{enumitem}

\graphicspath{{fig/}}
\usepackage{hyperref}
\usepackage{multirow,xcolor}
\hypersetup{
    colorlinks = true,
    linkcolor = blue,
    linkbordercolor = white,
    citecolor=blue,
    urlcolor=black
}
\usepackage[export]{adjustbox}
\usepackage{subfig,caption}
\newcommand{\abinitio}{\textit{ab initio }}

\newcommand{\bh}[1]{\hat{\mathbf{#1}}}

\newcommand{\Qb}{\mathbf{Q}}
\newcommand{\qb}{\mathbf{q}}
\newcommand{\Rb}{\mathbf{R}}
\newcommand{\Pb}{\mathbf{P}}

\newcommand{\db}{\mathbf{d}}
\newcommand{\cev}[1]{\reflectbox{\ensuremath{\vec{\reflectbox{\ensuremath{#1}}}}}}

\makeatletter
\let\cat@comma@active\@empty
\makeatother

\begin{document}
\title{A Quantum-Classical Liouville Formalism in a Preconditioned Basis and Its Connection with Phase-Space Surface Hopping}
\author{Yanze Wu}
\email{wuyanze@sas.upenn.edu}
\affiliation{Department of Chemistry, University of Pennsylvania, Philadelphia, Pennsylvania 19104, USA}
\author{Joseph E. Subotnik}
\email{subotnik@sas.upenn.edu}
\affiliation{Department of Chemistry, University of Pennsylvania, Philadelphia, Pennsylvania 19104, USA}
\date{\today}

\begin{abstract}
We revisit a recent proposal to model nonadiabatic problems with a complex-valued Hamiltonian through a phase-space surface hopping (PSSH) algorithm employing a pseudo-diabatic basis. Here, we show that such a pseudo-diabatic PSSH (PD-PSSH) ansatz is consistent with a quantum-classical Liouville equation (QCLE) that can be derived following a preconditioning process, and we demonstrate that a proper PD-PSSH algorithm is able to capture some geometric magnetic effects (whereas the standard FSSH approach cannot). We also find that a preconditioned QCLE can outperform the standard QCLE in certain cases, highlighting the fact that there is no unique QCLE. Lastly, we also point out that one can construct a mean-field Ehrenfest algorithm using a phase-space representation similar to what is done for PSSH. These findings would appear extremely helpful as far understanding and simulating nonadiabatic dynamics with complex-valued Hamiltonians and/or spin degeneracy.
\end{abstract}

\maketitle

\section{Introduction} \label{sec:introduction}

Over the last few decades, the quantum-classical Liouville equation (QCLE) has emerged as a crucial theoretical tool for understanding nonadiabatic dynamics and has functioned as a practical tool for high accuracy simulations for a small class of systems \cite{kapral1999Mixed,martens1997Semiclassicallimit,nielsen2000Mixed,donoso2000Semiclassical,kim2008Quantumclassical,kelly2012Mapping,shakib2014analysis}. Most importantly, the QCLE has been found to have a strong connection to a variety of surface hopping algorithms,\cite{martens2016Surface} especially with Tully's fewest switch surface hopping (FSSH)~\cite{tully1990Molecular}, one of the most popular nonadiabatic algorithms. With a decent decoherence algorithm and if momenta are relatively large, FSSH can capture many features predicted by the QCLE~\cite{subotnik2013Can,kapral2016Surface}.

Notably, one of the strengths of the QCLE is that the algorithm can capture (at least some) Berry force effects which might arise, for example, when one propagates nuclei over a potential energy surface generated with a Hamiltonian that lacks time reversal symmetry, i.e. in a magnetic field.\cite{wu2020Chemical,bian2021Modeling} In particular, if one projects all of the QCLE dynamics onto a mean-field electronic state, a Berry force does emerge.\cite{subotnik2019demonstration} That being said, Tully's original FSSH cannot recover any such Berry curvature effects at all~\cite{miao2019extension}.
%which might appear in systems with spin-orbit couplings or magnetic fields~\cite{wu2021Electronic,bian2021Modeling,chandran2022Electron,teh2021Antisymmetric}.
In fact, Tully's FSSH algorithm cannot readily treat complex-valued Hamiltonians in the first place: From a practical perspective, FSSH requires a real-valued direction for the derivative coupling vector $\mathbf{d}$ in order to perform momentum rescaling, but FSSH has a problematic (non-unique) gauge dependence in the context of a complex-valued Hamiltonian.
To that extent, recently our group has proposed to solve such a problem by using a phase-space surface hopping (PSSH) approach based on a phase transformed diabatic basis (called `pseudo-diabats') \cite{wu2022phasespace}. In our tests of model systems, such a pseudo-diabatic PSSH (PD-PSSH) guess can indeed model complex-valued Hamiltonians and incorporate Berry curvature effects while retaining much of the simplicity of standard FSSH. \cite{wu2022phasespace,bian2022Modeling}

Despite the practical successes demonstrated in Refs.~\cite{wu2022phasespace,bian2022Modeling}, the underlying theory behind the PD-PSSH algorithm has heretofore not been known. Whereas Subotnik\cite{subotnik2013Can} and Kapral\cite{kapral2016Surface} {\em et al} have been able to understand how the FSSH algorithm can be roughly mapped to the QCLE for a standard real-valued (non-degenerate) Hamiltonian, such a relationship between a PD-PSSH and the QCLE has not yet been demonstrated. Thus, with this background in mind, our goal here is to map a broad class of PSSH algorithms to the QCLE which, as we have noted above, does capture Berry effects. Importantly, however, our mapping below will connect a PD-PSSH approach not directly to the standard QCLE, but rather to a non-standard dressed QCLE, i.e. a QCLE where the Wignerization is performed in a certain basis (a basis of ``pseudo-diabats'') rather than a purely diabatic basis. One of the conclusions of the present manuscript is that, just as there many incarnations of surface hopping dynamics, there are also several valid, different incarnations of the QCLE.
In fact, in order to make our discussion of the QCLE as clear and comprehensive as possible, for future reference below, it will be essential to have very clear definitions and nomenclature as far as book-keeping the various QCLE and surface hopping algorithms that can be generated. These conventions are summarized in Table ~\ref{table:qcle}. 
\begin{table}[H]
\begin{center}
\begin{tabular}{c|c|c}
Basis for Wignerization & QCLE & Corresponding SH \\
\hline
diabats & standard (diabatic) QCLE (D-QCLE)\cite{kapral1999Mixed} Eq.~\eqref{eq:dqcle} & FSSH\cite{tully1990Molecular} \\
a preconditioned basis & preconditioned QCLE (P-QCLE) Eq.~\eqref{eq:pqcle} & PD-PSSH\cite{wu2022phasespace} \\
 adiabats & adiabatic-then-Wigner QCLE (A-QCLE)\cite{ryabinkin2014Analysis} Eq.~\eqref{eq:aqcle} & A-PSSH\cite{shenvi2009Phasespace}
\end{tabular} \caption{Our naming conventions for the QCLE and surface hopping (SH) algorithms. Each surface hopping algorithm is connected to the QCLE in the same row by a set of similar approximations.} \label{table:qcle}
\end{center} 
\end{table}

Returning to the central goal of this paper, i.e., mapping the PD-PSSH algorithm to the QCLE, our approach will be as follows:
\begin{enumerate}
\item We will begin by introducing the notion of a ``preconditioned'' QCLE (P-QCLE), where we transform the density matrix into a pseudo-diabatic basis (a ``preconditioning'' process) before implementing a Wignerization step. Over the previous two decades, several groups have transformed the density matrix to an adiabatic basis before Wignerization \cite{horenko2002Quantumclassical,donoso2000Semiclassical,ando2003Mixed} and found that reasonable results as compared with the has produced similar results as the standard QCLE\cite{donoso2000Semiclassical,ando2003Mixed}. However, more recently, Ryabinkin et al have shown that such an ``adiabatic-then-Wigner'' QCLE can cause problems when geometric phase effects are significant \cite{ryabinkin2014Analysis}, e.g. near conical intersections; thus, at present, the standard QCLE (or D-QCLE) is that equation that arises when the quantum-mechanical operators are Wignerized in a diabatic representation \cite{kapral1999Mixed}, a "Wigner then Adiabatic approach." 
For our present paper, we will argue that if we transform to a pseudo-diabatic basis that changes smoothly and slowly as a function of nuclear configuration, the problems highlighted in Ref.~\cite{ryabinkin2014Analysis} should not emerge. Moreover, we will show that a P-QCLE can actually have some advantages, e.g. the algorithm can indirectly include some elements of a diagonal Born-Oppenheimer corrections (DBOC) \cite{lengsfield1992Nonadiabatic} (which the standard QCLE does not include). See Sec.~\ref{sec:shenvi}.

\item We will demonstrate that the pseudo-diabatic PSSH (PD-PSSH) algorithm as proposed in Ref.~\cite{wu2022phasespace,bian2022Modeling} is consistent with a P-QCLE, much the same way as FSSH is consistent with the standard QCLE. We will show that Tully's FSSH misses a coherence term contained within the QCLE, which in turn arises from coherent interactions between the active diagonal state and adjacent off-diagonal states and is nonzero only when the derivative couplings are complex-valued. In the adiabatic limit, this term becomes effectively a Berry force. Luckily, by transforming to a real-valued representation (with real-valued derivative couplings), the PD-PSSH does not suffer from such a problem.

\item Lastly, we will demonstrate why, even for a system with time-reversal symmetry (for which the on-diagonal Berry curvature is zero), the PD-PSSH algorithm is able to recover exotic pseudo-magnetic fields. These wavepacket bending and reflecting effects can arise whenever one reaches a singlet-triplet crossing.
\end{enumerate}

Throughout this paper, we assume a hat symbol $\hat{O}$ represents a total nuclear-electronic quantum-mechanical operator, while a bold color symbol $\mathbf{A}$ (except $\mathbf{q}$, which represents the electronic degrees of freedom) represents a vector in the nuclear space. The vector dot $\mathbf{A}\cdot\mathbf{B}$ is always performed in the nuclear space, and if both are matrices, their matrix multiplication is also performed, i.e. $(\mathbf{A}\cdot\mathbf{B})_{jk}=A_{jl}^\alpha B_{lk}^\alpha$. The Einstein summation convention will also be used extensively for matrix and vector multiplications. For maximum clarity, we list our definition of indices in Table \ref{table:indices} and some of our symbols in Table \ref{table:symbol}.
\begin{table}[H]
\begin{center}
\begin{tabular}{c|l}
\hline
Indices & What they label \\ \hline
$\alpha,\beta$ & Nuclear degrees of freedom \\
$j_0,k_0,l_0$ & Diabats \\
$j,k,l$ & Pseudo-diabats \\
$m,n,s$ & Phase-space adiabats \\
\hline
\end{tabular} \caption{Definition of Indices}\label{table:indices}
\end{center}
\end{table}

\begin{table}[H]
\begin{center}
\begin{tabular}{c|l}
\hline
Symbol & Meaning \\ \hline
$\mathbf{Q}$ & Quantum mechanical nuclear position operator \\
$\mathbf{q}$ & Quantum mechanical electronic position operator \\
$\Rb,R_\alpha$ & Classical nuclear position \\
$\Pb,P_\alpha$ & Classical nuclear momentum \\
$\hat{H},\hat{H}_{jk}$ & Total quantum mechanical Hamiltonian \\
$H_W,H_{jk}^W$ & Wigner transform of $\hat{H}$ in a pseudo-diabatic basis \\
$E$ & Phase-space adiabatic energies, i.e., the eigenvalues of $H_W$ \\
$\hat{h}$ & Electronic quantum mechanical Hamiltonian \\
$h_W,h_{jk}^W $ & Wigner transform of $\hat{h}$ in a pseudo-diabatic basis \\
$h$ & Representation of $h_W$ in a phase-space adiabatic basis \\
$\hat{\rho}$ & Quantum mechanical density matrix \\
$\rho_W$ & Wigner transformed density matrix in a pseudo-diabatic basis \\
$\rho$ & Representation of $\rho_W$ in a phase-space adiabatic basis \\
$\hat{\mathbf{D}}$ & Quantum mechanical derivative coupling operator between pseudo-diabats \\
$\mathbf{D}_W,D_\alpha^W,D^{W\alpha}_{jk}$ & Wigner transform of $\hat{\mathbf{D}}$ in a pseudo-diabatic basis \\
$\mathbf{D},D^\alpha_{mn}$ & Representation of $\mathbf{D}_W$ in a phase-space adiabatic basis \\
$\db,d_{mn}^\alpha$ & Partial derivative coupling between phase-space adiabats with respect to $\Rb$ \\
$\bm{\tau},\tau_{mn}^\alpha$ & Derivative coupling of the phase-space adiabats with respect to $\mathbf{P}$ \\
$\db_A$ & Derivative coupling between (position-space) adiabats \\
\hline
\end{tabular} \caption{List of Symbols and Definition}\label{table:symbol}
\end{center}
\end{table}

\section{The Preconditioned Quantum-Classical Liouville Equation}

In our work below, it will be essential to transform the quantum-mechanical Liouville equation into a pseudo-diabatic basis before we Wignerize and derive a QCLE. To that extent, it is now appropriate to recapitulate a great deal of the QCLE formalism in an arbitrary basis (basically following Ref.~\cite{donoso2000Semiclassical,ando2003Mixed,ryabinkin2014Analysis}) so that we can easily analyze the resulting coupled nuclear-electronic equation. We emphasize that in the end, our choice of pseudo-diabatic representation should change smoothly as a function of nuclear coordinates. In such a case, even though the resulting QCLE may not be exact for the spin-boson model\cite{mackernan2002Surfacehopping}, we do not anticipate finding the same large and spurious errors as we would find if we were to follow an adiabatic-then-Wigner QCLE algorithm around a conical intersection. Moreover, as we show numerically in Sec.~\ref{sec:shenvi}, for certain model Hamiltonians, the resulting non-standard QCLE can recover some features that the standard QCLE simply cannot.

\subsection{QCLE in an Arbitrary Basis} \label{sec:qcle1}
To begin with, let us consider a general nonadiabatic Hamiltonian with nuclear degrees of freedom $\Qb$ and electronic degrees of freedom $\qb$:
\begin{align}
    \hat{H}(\Qb,\qb) = -\frac{\nabla_Q^2}{2M}-\frac{\nabla_q^2}{2m}+\hat{h}(\Qb,\qb)
\end{align}
The corresponding quantum Liouville equation reads
\begin{align} \label{eq.qle}
    \pdv{\hat{\rho} (\Qb,\qb)}{t} = -\frac{i}{\hbar}\left[\hat{H}(\Qb,\qb),\hat{\rho} (\Qb,\qb)\right]
\end{align}

%In the standard approach towards deriving the QCLE, Eq.~\eqref{eq.qle} is Wigner transformed before projected to a specific basis.
%It has previously been shown that, when one works with such a QCLE, the final result is exact for a linear-coupled spin-boson system, and that Berry geometric phase can be recovered by the resulting dynamics as well. Previous work has also shown that this approach performs better than the alternative, AW, whereby ...\cite{XXX}
In contrast to the standard approach where Eq.~\eqref{eq.qle} is Wigner transformed and then projected into a specific basis, we will first apply a preconditioning by transforming Eq.~\eqref{eq.qle} into a new electronic basis $\{\ket{j(\Qb)}\}$ and then we apply a Wigner transformation. We will call this initial basis a ``pseudo-diabatic'' basis; in general, we expect this basis to be neither strictly diabatic nor adiabatic. See below. For simplicity, we assume the pseudo-diabats are single-valued, orthonormal and complete. 
Note that mathematically, there are no additional restrictions for our pseudo-diabats, however in practice, the choice of pseudo-diabats will strongly affect our interpretation and simulations; see Sec.~\ref{sec:pseudodiabat}. 

In a pseudo-diabatic basis, Eq.~\eqref{eq.qle} reads
\begin{align} \label{eq:qle2}
    \pdv{\hat{\rho}_{jk}}{t} = -\frac{i}{\hbar}(\hat{H}_{jl}\hat{\rho}_{lk}-\hat{\rho}_{jl}\hat{H}_{lk})
\end{align}
where $\hat{\rho}_{jk}(\Qb)=\mel{j(\Qb)}{\hat{\rho}}{k(\Qb)}$ and $\hat{H}_{jk}(\Qb)=\mel{j(\Qb)}{\hat{H}}{k(\Qb)}$ are the projected operators. From Born-Oppenheimer theory, the Hamiltonian element $\hat{H}_{jk}$ is
\begin{align} \label{eq:hqm}
   \hat{H}_{jk}(\Qb) = \hat{h}_{jk}(\Qb) -\frac{\hbar^2}{2M}{\left(\nabla_Q+\bh{D}\right)}^2_{jk}
\end{align}
where $\bh{D}_{jk}=\braket{j}{\nabla_Qk}$ is the derivative coupling.

At this point, we apply a partial Wigner transform on the nuclear coordinates. For an arbitrary operator $\hat{O}$ (which could be, e.g., $\hat{\rho},\hat{h}$), we define
\begin{align}
    O^W_{jk}(\Rb,\Pb)=\int{\mel{\Rb+\frac{\mathbf{y}}2}{\hat{O}_{jk}}{\Rb-\frac{\mathbf{y}}2}e^{-\frac{i}{\hbar}{\mathbf{y}\cdot\Pb}}d\mathbf{y}}
\end{align}
As pointed in Ref.~\cite{ryabinkin2014Analysis}, if the pseudo-diabatic basis is ``double-valued'' (e.g. the adiabats near a conical intersection), the Wigner transformation will face some boundary-value problems. However, as mentioned above, we assume our pseudo-diabats do not suffer from this problem, which is indeed true for our choices described in Ref.~\cite{wu2022phasespace} or in Sec.~\ref{sec:pseudodiabat}. After the Wigner transform, Eq.~\eqref{eq:qle2} becomes
\begin{align} \label{eq:qclefull}
    \pdv{\rho^W_{jk}}{t} =-\frac{i}{\hbar}\left(H^W_{jl}e^{-i\hbar\hat{\Lambda}/2}\rho^W_{lk}-\rho^W_{jl}e^{-i\hbar\hat{\Lambda}/2}H^W_{lk}\right)
\end{align}
where $\hat{\Lambda} = \cev{\nabla}_P\cdot\vec{\nabla}_R-\cev{\nabla}_R\cdot\vec{\nabla}_P$ is the Poisson bracket operator. Eq.~\eqref{eq:qclefull} is exact.

As in the standard QCLE derivation, we now take only the zeroth and first order terms in $\hbar$ for the Taylor expansion of the exponential term, i.e. $e^{-i\hbar\hat{\Lambda}/2}\approx 1-i\hbar\hat{\Lambda}/2$. Now Eq.~\eqref{eq:qclefull} reads
\begin{equation} \label{eq:qcle}
\begin{split}
    \pdv{\rho^W_{jk}}{t} = &-\frac{i}{\hbar}\left[H_{jl}^W\rho_{lk}^W-\rho_{jl}^WH_{lk}^W\right]-\frac{1}{2}\left(\pdv{H_{jl}^W}{P_\alpha}\pdv{\rho_{lk}^W}{R_\alpha}+\pdv{H_{lk}^W}{P_\alpha}\pdv{\rho_{jl}^W}{R_\alpha}\right) \\ &+\frac{1}{2}\left(\pdv{H_{jl}^W}{R_\alpha}\pdv{\rho_{lk}^W}{P_\alpha}+\pdv{H_{lk}^W}{R_\alpha}\pdv{\rho_{jl}^W}{P_\alpha}\right) 
\end{split}
\end{equation}

%For the second approximation, we drop the divergence part in the second derivative coupling matrix, i.e. $G_{jk}^W=-\nabla_Q\cdot\mathbf{D}_{jk}^W+\mathbf{D}_{jl}^W\cdot \mathbf{D}_{lk}^W\approx \mathbf{D}_{jl}^W\cdot \mathbf{D}_{lk}^W$. 
The Wigner transform of the Hamiltonian \eqref{eq:hqm} reads \cite{wigner}:
\begin{align} \label{eq:hsc}
    H^W_{jk}(\Rb,\Pb)=h^W_{jk}(\Rb)+\frac{{(\Pb-i\hbar\mathbf{D}_W(\Rb))}^2_{jk}}{2M}
\end{align}
where $\Pb$ is an identity matrix of the nuclear momenta. Plugging Eq.~\eqref{eq:hsc} into Eq.~\eqref{eq:qcle}, and writing the equation in matrix form, we arrive at the preconditioned QCLE:
\begin{align}
    \pdv{\rho_W}{t}=-\frac{i}{\hbar}\left[H_W,\rho_W\right] -\frac{1}{2}\left[\frac{P_\alpha-i\hbar D_\alpha^W}{M},\pdv{\rho_W}{R_\alpha}\right]_+ - \frac{1}{2}\left[F_\alpha^W,\pdv{\rho_W}{P_\alpha}\right]_+
    \label{eq:pqcle}
\end{align}
where $([a,b]_+)_{jk}=a_{jl}b_{lk}+b_{jl}a_{lk}$ is the matrix anticommutator, and $F_\alpha^W\equiv-\partial H_W/\partial R_\alpha$ is the nuclear force.

Compared to the standard QCLE, the P-QCLE allows for one more degree of freedom in any optimization process, namely one has the freedom to choose a basis of pseudo-diabats so as to best match exact quantum mechanics: If we choose the pseudo-diabats to be a set of pure diabats, $\mathbf{D}$ would be zero and Eq.~\eqref{eq:pqcle} would be equivalent to the standard QCLE\cite{kapral1999Mixed}:
\begin{align}
    \pdv{\rho_W}{t}=-\frac{i}{\hbar}\left[h_W,\rho_W\right] -\frac{P_\alpha}{M}\pdv{\rho_W}{R_\alpha} - \frac{1}{2}\left[F_\alpha^W,\pdv{\rho_W}{P_\alpha}\right]_+
    \label{eq:dqcle}
\end{align}
By contrast, if we choose the pseudo-diabats to be the same as a set of adiabats, then $\mathbf{D}$ would equal to adiabatic derivative coupling $\db_A$ and Eq.~\eqref{eq:pqcle} would be equivalent to the adiabatic-then-Wigner QCLE\cite{ryabinkin2014Analysis} (here $H_A$ and $F_A$ are the analogues in the adiabatic basis of $H_W$ and $F_W$ in a diabatic basis):
\begin{align}
    \pdv{\rho_W}{t}=-\frac{i}{\hbar}\left[H_A,\rho_W\right] -\frac{1}{2}\left[\frac{P_\alpha-i\hbar d_\alpha^A}{M},\pdv{\rho_W}{R_\alpha}\right]_+ - \frac{1}{2}\left[F_\alpha^A,\pdv{\rho_W}{P_\alpha}\right]_+
    \label{eq:aqcle}
\end{align}
We will argue below that, in order to construct the most accurate surface hopping approach, the optimal pseudo-diabatic basis should be neither strictly adiabatic or strictly diabatic.

\subsection{Representation in a Basis of Phase-Space Adiabats}

Following Kapral's approach to surface hopping dynamics\cite{kapral2016Surface}, the next step is to represent Eq.~\eqref{eq:pqcle} in the eigenbasis of $H_W$ (known as the phase-space adiabats~\cite{wu2022phasespace,shenvi2009Phasespace}). By multiplying phase-space adiabats $\ket{m}$, $\ket{n}$ onto the two sides of Eq.~\eqref{eq:pqcle}, we find
\begin{align}
    \pdv{\rho_{mn}}{t} \equiv \mel{m}{\pdv{\rho_W}{t}}{n} =&-\frac{i}{\hbar}\mel{m}{\left[H_W,\rho_W\right]}{n} \nonumber\\ &-\frac{1}{2}\mel{m}{\left[\frac{P_\alpha-i\hbar D^W_\alpha}{M},\pdv{\rho_W}{R_\alpha}\right]_+}{n} + \frac{1}{2}\mel{m}{\left[F_\alpha^W,\pdv{\rho_W}{P_\alpha}\right]_+}{n} 
    \label{eq:pqcle1}
\end{align}
To simplify Eq.~\eqref{eq:pqcle1}, we utilize the following identities or definitions:
\begin{align}
    D_{mn}^\alpha \equiv \mel{m}{D_\alpha^W}{n} &= \braket{m}{j}D^{W\alpha}_{jk}\braket{k}{n} \label{eq:dmn} \\
    \pdv{\rho_{mn}}{P_\alpha} \equiv \mel{m}{\pdv{\rho_W}{P_\alpha}}{n} &=
    \pdv{\rho_{mn}^W}{P_\alpha}-\braket{\pdv{m}{P_\alpha}}{s}\rho_{sn}-\rho_{ms}\braket{s}{\pdv{n}{P_\alpha}} \nonumber\\
    &=\pdv{\rho_{mn}^W}{P_\alpha}+\braket{m}{\pdv{s}{P_\alpha}}\rho_{sn}-\rho_{ms}\braket{s}{\pdv{n}{P_\alpha}} \nonumber\\ 
    &\equiv\pdv{\rho_{mn}^W}{P_\alpha}+\tau^\alpha_{ms}\rho_{sn}-\rho_{ms}\tau^\alpha_{sn}
    \label{eq:drhodp} \\
    \pdv{\rho_{mn}}{R_\alpha} \equiv \mel{m}{\pdv{\rho_W}{R_\alpha}}{n} &=\braket{m}{j}\pdv{\rho^W_{jk}}{R_\alpha}\braket{k}{n} \nonumber\\ &=
    \pdv{\rho_{mn}^W}{R_\alpha}+\braket{m}{j}\pdv{\braket{j}{s}}{R_\alpha}\rho_{sn}-\braket{s}{k}\pdv{\braket{k}{n}}{R_\alpha}\rho_{ms} \nonumber\\
    &\equiv\pdv{\rho_{mn}^W}{R_\alpha}+d^\alpha_{ms}\rho_{sn}-\rho_{ms}d^\alpha_{sn}
    \label{eq:drhodr}
\end{align}
Here in Eq.~\eqref{eq:drhodp} and \eqref{eq:drhodr}, we have defined $d_{mn}^\alpha=\braket{m}{j}\pdv{\braket{j}{n}}{R_\alpha}$ and $\tau_{mn}^\alpha =\braket{m}{\pdv{n}{P_\alpha}}$; these quantities are the derivative couplings of phase-space adiabats with respect to $\Rb$ and $\Pb$.

Finally, by plugging Eqs.~\eqref{eq:dmn}-\eqref{eq:drhodr} into Eq.~\eqref{eq:pqcle1} for all possible $m$ and $n$, and after a little bit of manipulation, we can transform Eq.~\eqref{eq:pqcle1} into a matrix form:
\begin{align}
   \pdv{\rho}{t}=&-\frac{i}{\hbar}[E,\rho]-\frac{1}{2}\left[\frac{P_\alpha-i\hbar D_\alpha}{M},\pdv{\rho}{R_\alpha}\right]_+-\frac{1}{2}\left[\frac{P_\alpha-i\hbar D_\alpha}{M},[d_\alpha,\rho]\right]_+ \nonumber\\
    &-\frac{1}{2}{\left[F_\alpha,\pdv{\rho}{P_\alpha}\right]}_+-\frac{1}{2}{\left[F_\alpha,[\tau_\alpha,\rho]\right]}_+ \nonumber\\
    =&-\frac{i}{\hbar}\left[E-i\hbar\frac{P_\alpha d_\alpha}{M},\rho\right] -\frac{1}{2}\left[\frac{P_\alpha-i\hbar D_\alpha}{M},\pdv{\rho}{R_\alpha}\right]_+ -\frac{1}{2}\left[F_\alpha,\pdv{\rho}{P_\alpha}\right]_+ \nonumber\\
    &+\frac{i\hbar}{2M}{\left[D_\alpha,[d_\alpha,\rho]\right]}_+-\frac{1}{2}{\left[F_\alpha,[\tau_\alpha,\rho]\right]}_+
     \label{eq:pqclead}
\end{align}
Note that all the operators in Eq.~\eqref{eq:pqclead} are defined in a basis of the phase-space adiabats $m,n$, in contrast to the operators in Eq.~\eqref{eq:pqcle} which are defined in a basis of pseudo-diabats $j,k$. In short, Eq.~\eqref{eq:pqclead} is the P-QCLE represented in the phase-space adiabatic basis. Below, this equation, will be used below to make connections with PD-PSSH algorithms.

%Compared to the standard WA-QCLE , Eq.~\eqref{eq:pqclead} gets two extra nested commutators and additional $-i\hbar D_\alpha$ in the velocity part. 

%Finally, to make connections with surface hopping, we employ the following two assumptions:
%\begin{align}
%    \left[D_\alpha,[d_\alpha,\rho]\right]_+&= [[D_\alpha,d_\alpha]_+, \rho] + [[\rho,D_\alpha],d_\alpha]_+ \approx [[D_\alpha,d_\alpha]_+, \rho] \\
%    \left[F_\alpha,[\tau_\alpha,\rho]\right]_+ &=[[F_\alpha, \tau_\alpha]_+, \rho] + [[\rho,F_\alpha],\tau_\alpha]_+ \approx [[F_\alpha, \tau_\alpha]_+, \rho] \label{eq:defu}
%\end{align}
%The physical meaning is that we are assuming only the interaction between diagonal part $\rho$ and $D$ or $F$ are important. By defining the velocity operator $v_\alpha=(P_\alpha-i\hbar D_\alpha)/M$, Eq.~\eqref{eq:pqclead} becomes
%\begin{align}
%   \pdv{\rho}{t}=&-\frac{i}{\hbar}\left[E-\frac{i\hbar}{2}[v_\alpha,d_\alpha]_+-\frac{i\hbar}{2}[ F_\alpha,\tau_\alpha]_+,\rho\right]
%   -\frac{1}{2}\left[v_\alpha,\pdv{\rho}{R_\alpha}\right]_+  -\frac{1}{2}\left[F_\alpha,\pdv{\rho}{P_\alpha}\right]_+ \label{eq:qcleps}
%\end{align}

\section{Phase-space Surface Hopping} \label{sec:pssh}
Recently, our research group has suggested a pseudo-diabatic phase-space surface hopping (PD-PSSH) for solving nonadiabatic problems with complex-valued Hamiltonians and/or degeneracy as would be standard in the presence of several spin states. This algorithm can be described as follows: A swarm of trajectories are generated according to the Wigner distribution, each assigned with an active phase-space adiabat $\lambda$, an electronic density matrix $\sigma$, position $\Rb$ and momentum $\Pb$. These quantities are then propagated according to the following equations of motion:
\begin{align}
    \dot{\sigma}&= -\frac{i}{\hbar}\left[E(\Rb,\Pb)-i\hbar\dot{\Rb}\cdot\db-i\hbar\dot{\Pb}\cdot\bm{\tau},\sigma\right] \label{eq:shsigma} \\
    \dot{\Rb}&=\mel{\lambda}{\nabla_PE_\lambda}{\lambda}=\frac{\Pb-i\hbar\mathbf{D}_{\lambda\lambda}}{M} \label{eq:shr} \\
    \dot{\Pb}&=-\mel{\lambda}{\nabla_RE_\lambda}{\lambda}=\mathbf{F}_{\lambda\lambda} \label{eq:shp}
\end{align}
The trajectory's active surface $\lambda$ is changed (i.e. we ``hop'') to match the population dynamics $\sigma$. The hopping rate from surface $m\to n$ is given by
\begin{align}
    g_{m\to n} = \max\left[ 2\Re\left(\frac{\sigma_{nm}(\dot{\Rb}\cdot\db_{mn}+\dot{\Pb}\cdot\bm{\tau}_{mn})}{\sigma_{mm}}\right), 0\right]
    \label{eq:shhop}
\end{align}
When a hop is attempted, the trajectory's momentum is adjusted along the $\db_{nm}$ direction in order to maintain energy conservation, i.e. $\Delta \Pb$ is chosen to satisfy $E_m(\Rb,\Pb+\mathbf{d}_{nm}\Delta \Pb)=E_n(\Rb,\Pb)$. We will show in Sec.~\ref{sec:pseudodiabat} that our choices of the pseudo-diabats will make $\db_{nm}$ real-valued, therefore we will not encounter any ambiguities about the rescaling direction.

The only difference between PSSH and Tully's FSSH is the $\Pb$-dependence of the Hamiltonian, i.e. we have $E(\Rb,\Pb)$ instead of $E(\Rb)$. Thus, one might reasonably expect that the connection between FSSH and standard QCLE\cite{subotnik2013Can,kapral2016Surface} should also extend to a connection between PSSH and the P-QCLE. In this paper, to establish such a connection, we will follow the frameworks of both Subotnik \cite{subotnik2013Can} and Kapral \cite{kapral2016Surface} {\em et al}. As a brief review, Tully-style surface hopping dynamics can be connected to the standard QCLE if one makes the following approximations:
\begin{enumerate}
\item Unique-trajectory assumption: We assume that only one single trajectory reaches a given point in phase space ($\Rb$,$\Pb$) and a given surface $\lambda$ at time $t$. Surface hopping then estimates the density matrix by
\begin{align}
   \rho^{\text{SH}}_{nn}(\Rb,\Pb) \equiv \frac{1}{N_{\text{traj}}}\sum_{\text{traj}}{\delta_{\lambda(t)n}\delta(\Rb-\Rb(t),\Pb-\Pb(t))}
\end{align}
\item Large velocity assumption: We assume that the momentum adjustment $\Delta\Pb$ during a hop can be treated as a small variable in the expansion. Depending on the context, this assumption means we can ignore all $O({(\Delta P/P)}^2)$ and $O({(\Delta P\pdv{\sigma}{P})}^2)$ terms (where $\sigma$ is the electronic density matrix). Moreover, we assume that we can always find a real-valued solution for the momentum rescaling; there is currently no means to map Tully's frustrated hops to the QCLE.
\item Decoherence assumption: We must assume that the off-diagonal density matrix element $\sigma_{nm}$ is an accurate estimate of the partial Wigner density matrix $\rho_{nm}$, i.e. 
\begin{align}
    \frac{1}{N_{\text{traj}}}\sum_{\text{traj}}{\sigma_{nm}(t)\delta(\Rb-\Rb(t),\Pb-\Pb(t))} \approx \rho_{nm}(\Rb,\Pb,t)
    \label{eq:rhooffdiag}
\end{align}
Note that for Tully's original FSSH, Eq.~\eqref{eq:rhooffdiag} is not satisfied, which yields the so-called decoherence problem. \cite{subotnik2013Can,subotnik2016Understanding} Below, we will not treat decoherence and we will assume that this problem has been fixed by some decoherence patch or another. We will provide a short discussion about the applicability of the AFSSH decoherence scheme \cite{subotnik2011new,jain2016Efficient} for PSSH in Sec.~\ref{sec:dyn}.
\end{enumerate}

Now, in order to map the PD-PSSH to the P-QCLE, we will adopt all of the approximations above, and we will also need to make two additional assumptions:
\begin{enumerate}
\setcounter{enumi}{3}
\item When propagating the electronic amplitude, we will estimate that the two nested commutators $[D_\alpha,[d_\alpha,\rho]]_+$ and $[F_\alpha,[\tau_\alpha,\rho]]_+$ in Eq.~\eqref{eq:pqclead} can be approximated with pure commutators whose matrix elements can be taken from a local trajectory. More precisely, we will approximate
\begin{align}
    -\left[\frac{P_\alpha d_\alpha}{M}, \rho\right] + \frac{i\hbar}{2M}[D_\alpha,[d_\alpha,\rho]]_+ &\approx -\left[\dot{R}^\alpha_{\text{SH}} d_\alpha, \rho\right] \label{eq:commapprox1} \\
    -\frac{1}{2}[F_\alpha,[\tau_\alpha,\rho]]_+ &\approx -\left[\dot{P}^\alpha_{\text{SH}}\tau_\alpha, \rho\right] \label{eq:commapprox2}
\end{align}
where $\dot{R}_{\text{SH}}$ and $\dot{P}_{\text{SH}}$ are properties of the trajectory estimated by Eqs.~\eqref{eq:shr} and \eqref{eq:shp}. We will discuss these approximation in more details in Sec.~\ref{sec:ele}.
\item When calculating the momentum rescaling, we will assume that the momentum is large enough, so that we may simply drop all of the nested commutator terms ($[D_\alpha,[d_\alpha,\rho]]_+$ and $[F_\alpha,[\tau_\alpha,\rho]]_+$) entirely.
\end{enumerate}

\subsection{The Structure of the QCLE}

\newcommand{\Jnn}[1]{\ensuremath{J_{nn'}^{\text{#1}}}}
\newcommand{\Jnnmm}[1]{\ensuremath{J_{nn',mm'}^{\text{#1}}}}

Keeping in mind the approximations that we will need to make in order to show the consistency of PD-PSSH and the P-QCLE, we will begin by separating the full P-QCLE (Eq.~\eqref{eq:pqclead}) into three parts: the electronic part $J^{\text{ele}}$, the off-diagonal force part $J^{\text{off}}$ and the dynamics part $J^{\text{dyn}}$:
\begin{align}
    \pdv{\rho_{nn'}}{t} &= \left(\Jnnmm{ele}+\Jnnmm{off}\right)\rho_{mm'}+\Jnn{dyn}\rho_{nn'}
    \label{eq:liouville}
\end{align}
where
\begin{align}
\begin{split}
    \Jnnmm{ele} =& -\frac{i}{\hbar}(E_n-E_{n'})\delta_{nm}\delta_{n'm'} -d_{nm}^\alpha\frac{P_\alpha}{M}\delta_{n'm'} +d_{m'n'}^\alpha \frac{P_\alpha}{M}\delta_{nm} \\
    &+\frac{i\hbar}{2M}\left(D^\alpha_{ns}(d^\alpha_{sm}\delta_{m'n'}-d^\alpha_{m'n'}\delta_{sm}) + D^\alpha_{sn'}(d^\alpha_{nm}\delta_{m's}-d^\alpha_{m's}\delta_{nm})\right) \\
    &-\frac{1}{2}\left(F^\alpha_{ns}(\tau^\alpha_{sm}\delta_{m'n'}-\tau^\alpha_{m'n'}\delta_{sm}) + F_{sn'}^\alpha(\tau^\alpha_{nm}\delta_{m's}-\tau^\alpha_{m's}\delta_{nm})\right)
\end{split}\label{eq:Lele} \\
\begin{split}
    \Jnnmm{off} =& \left(-\frac{1}{2}F_{nm}^\alpha{\pdv{}{P_\alpha}}(1-\delta_{nm}) +\frac{i\hbar}{2M}D_{nm}^\alpha\pdv{}{R_\alpha}\right)\delta_{n'm'} \\
    &+\left(-\frac{1}{2}F_{m'n'}^\alpha{\pdv{}{P_\alpha}}(1-\delta_{n'm'})+\frac{i\hbar}{2M}D_{m'n'}^\alpha\pdv{}{R_\alpha}\right)\delta_{nm} 
\end{split} \label{eq:Loff} \\
    \Jnn{dyn} =& -\left(\frac{P}{M}-i\hbar\frac{D_{nn}^\alpha+D_{n'n'}^\alpha}{2M}\right)\pdv{}{R_\alpha}-\frac{1}{2}(F^\alpha_{nn}+F^\alpha_{n'n'})\pdv{}{P_\alpha} \label{eq:Ldyn}
\end{align}
We will investigate the consistency of each component in the following sections.

\subsection{Consistency of the Electronic Propagation} \label{sec:ele}

The full electronic propagator $J^{\text{ele}}$ consists of two nested commutator terms, which cannot be simply handled by surface hopping without an expensive calculation of the explicit derivative couplings and forces. Therefore, we adopt the approximations in Eqs.~\eqref{eq:commapprox1} and \eqref{eq:commapprox2} in order to propagate the electronic amplitudes. The final $J^{\text{ele}}$ is then taken to be of the form:
\begin{align}
\begin{split}
    \Jnnmm{ele,SH} =& -\frac{i}{\hbar}(E_n-E_{n'})\delta_{nm}\delta_{n'm'} + \left(-d_{nm}^\alpha \dot{R}_\alpha^{\text{SH}}-\tau_{nm}^\alpha\dot{P}_\alpha^{\text{SH}}\right)\delta_{n'm'} \\
    &-\left(-d_{m'n'}^\alpha\dot{R}_\alpha^{\text{SH}}-\tau_{m'n'}^\alpha\dot{P}_\alpha^{\text{SH}}\right)\delta_{nm} 
\end{split}\label{eq:Lelesh}
\end{align}
The equation $\partial\rho_{nn'}/\partial t = \Jnnmm{ele,SH}\rho_{mm'}$ is exactly equivalent to the time-dependent Schrodinger equation~\eqref{eq:shsigma} in a basis that depends on both $\Rb$ and $\Pb$. The use of such a Schrodinger equation to propagate the electronic amplitudes is standard with surface hopping.

At this point, let us analyze the approximations made in Eqs.~\eqref{eq:commapprox1} and \eqref{eq:commapprox2} in more detail. For a trajectory on active surface $\lambda$, we have $\dot{R}^\alpha_{\text{SH}}=(P_\alpha - i\hbar D^\alpha_{\lambda\lambda})/M$ and $\dot{P}^\alpha_{\text{SH}}=F^\alpha_{\lambda\lambda}$. If we plug these identities into Eq.~\eqref{eq:Lelesh}, and subtract the result from Eq.~\eqref{eq:Lele}, we find
\begin{align}
\begin{split}
    \Jnnmm{ele} - \Jnnmm{ele,SH} =& \frac{i\hbar}{2M}\Big((D^\alpha_{ns}-D^\alpha_{\lambda\lambda}\delta_{ns})(d^\alpha_{sm}\delta_{m'n'}-d^\alpha_{m'n'}\delta_{sm}) \\
    &\qquad+(D^\alpha_{sn'}-D^\alpha_{\lambda\lambda}\delta_{sn'})(d^\alpha_{nm}\delta_{m's}-d^\alpha_{m's}\delta_{nm})\Big) \\
    &-\frac{1}{2}\Big((F^\alpha_{ns}-F^\alpha_{\lambda\lambda}\delta_{ns})(\tau^\alpha_{sm}\delta_{m'n'}-\tau^\alpha_{m'n'}\delta_{sm}) \\
    &\qquad+(F_{sn'}^\alpha-F^\alpha_{\lambda\lambda}\delta_{sn'})(\tau^\alpha_{nm}\delta_{m's}-\tau^\alpha_{m's}\delta_{nm})\Big)
\end{split} \label{eq:Lelediff}
\end{align}
By using the Hellman-Feynman theorem,
\begin{align}
    \tau_{mn}^\alpha &=\frac{\mel{m}{\partial H/\partial P_\alpha}{n}}{E_n-E_m} = -\frac{i\hbar}{M}\frac{D_{mn}^\alpha}{E_n-E_m}
\end{align}
we may further simplify the $\tau$ terms:
\begin{align}
\begin{split}
    \Jnnmm{ele} - \Jnnmm{ele,SH} =& \frac{i\hbar}{2M}\Big((D^\alpha_{ns}-D^\alpha_{\lambda\lambda}\delta_{ns})(d^\alpha_{sm}\delta_{m'n'}-d^\alpha_{m'n'}\delta_{sm}) \\
    &\qquad+(D^\alpha_{sn'}-D^\alpha_{\lambda\lambda}\delta_{sn'})(d^\alpha_{nm}\delta_{m's}-d^\alpha_{m's}\delta_{nm})\Big) \\
    &+\frac{i\hbar}{2M}\left(F^\alpha_{ns}-F^\alpha_{\lambda\lambda}\delta_{ns}\right)\left(\frac{D^\alpha_{sm}}{E_m-E_s}\delta_{m'n'}-\frac{D^\alpha_{m'n'}}{E_{n'}-E_{m'}}\delta_{sm}\right) \\
    &+\frac{i\hbar}{2M}\left(F_{sn'}^\alpha-F^\alpha_{\lambda\lambda}\delta_{sn'}\right)\left(\frac{D^\alpha_{nm}}{E_m-E_n}\delta_{m's}-\frac{D^\alpha_{m's}}{E_s-E_{m'}}\delta_{nm}\right) 
\end{split} \label{eq:Lelediff2}
\end{align}
Thus, we note that {\em all} of the terms in Eq.~\eqref{eq:Lelediff} are of order $\hbar$ and are expected to be relatively small if the trajectory has a large momentum and the $\mathbf{P}\cdot\mathbf{d}/M$ terms are dominant. Note that, within standard FSSH, there are similarly no terms proportional to $\hbar$ in the electronic Schrodinger equation.

\subsection{Consistency of Momentum Rescaling and Berry Force Effects} \label{sec:hop}

\newcommand{\drhodthop}{\ensuremath{\left.\pdv{\rho_{nn}(t)}{t}\right\vert_{\text{hop}}}}
\newcommand{\drhodtcoh}{\ensuremath{\left.\pdv{\rho_{nn}(t)}{t}\right\vert_{\text{coh}}}}
\newcommand{\Jdyn}[1]{\ensuremath{J^{\text{dyn}}_{#1}}}
\newcommand{\Jhop}[1]{\ensuremath{J^{\text{hop}}_{#1}}}

According to Ref.~\cite{kapral2016Surface}, momentum rescaling arises from the consecutive applications of the Liouvillian. In particular, the key terms for hopping and momentum rescaling are the off-diagonal terms in $J^{\text{ele}}$ and $J^{\text{off}}$. For the rest of the derivation, it will be convenient to define a hopping propagator as the sum of the off-diagonal terms in $J^{\text{ele}}$ and $J^{\text{off}}$ in Eqs.~\eqref{eq:Lele} and \eqref{eq:Loff} (again ignoring the commutators $[D_\alpha,[d_\alpha,\rho]]_+$ and $[F_\alpha,[\tau_\alpha,\rho]]_+$):\cite{commutator}
\begin{align}
\begin{split}
    \Jnnmm{hop} = & -d_{nm}^\alpha\frac{P_\alpha}{M}\delta_{n'm'}(1-\delta_{nm})+d_{m'n'}^\alpha \frac{P_\alpha}{M}\delta_{nm}(1-\delta_{n'm'}) \\ &-\frac{1}{2}F_{nm}^\alpha\pdv{}{P_\alpha}\delta_{n'm'}(1-\delta_{nm})-\frac{1}{2}F_{m'n'}^\alpha\pdv{}{P_\alpha}\delta_{nm}(1-\delta_{n'm'})
\end{split} \label{eq:Lhop}
\end{align}
Note that through $J^{\text{hop}}$, a diagonal element $\rho_{nn}$ interacts only with an off-diagonal element, while an off-diagonal element talks to both off-diagonal and diagonal elements. This property is important for our perturbation expansion below. %Note that a feature of $\Jc^{\text{res}}$ is that a diagonal element $\rho_{nn}$ interacts only with an off-diagonal element, while an off-diagonal element talks to both off-diagonal and diagonal elements. 

%To proceed with our derivation of the P-QCLE, we assume the momentum is large enough such that we can entirely drop the two nested commutator terms in Eq.~\eqref{eq:Lele}; XXX as just discussed, these terms are both of order $\hbar$. 
 
At this point, Eq.~\eqref{eq:liouville} becomes
\begin{align}
    \pdv{\rho_{nn'}}{t} &\approx \left(-\frac{i}{\hbar}(E_n-E_{n'})-(\db_{nn}-\db_{n'n'})\cdot\frac{\Pb}{M} + \Jnn{dyn}\right)\rho_{nn'} + \Jnnmm{hop}\rho_{mm'} \nonumber\\ 
    &\equiv \left(\Jnn{dyn}-i\tilde{\omega}_{nn'}\right)\rho_{nn'} + \Jnnmm{hop}\rho_{mm'} \label{eq:liouvilleapprox}
\end{align}
Here we define $\tilde{\omega}_{nn'}\equiv(E_n-E_{n'})/\hbar-i(\db_{nn}-\db_{n'n'})\cdot\Pb/M$. In Eq.~\eqref{eq:liouvilleapprox}, $(\Jnn{dyn}-i\tilde{\omega}_{nn'})$ consists of all diagonal interactions and $\Jnnmm{hop}$ consists of all off-diagonal interactions.
Next, by treating $\Jnnmm{hop}$ as a perturbation and $(\Jnn{dyn}-i\tilde{\omega}_{nn'})$ as the unperturbed part, we can expand the equation of motion for $\rho_{nn}$ as a time-dependent perturbation up to second order in $J^{\text{hop}}$ (and reduce the right hand side to a function of the diagonal elements or $\rho$ [albeit with memory]):
\begin{align}
    \pdv{\rho_{nn}(t)}{t} &\approx \Jdyn{nn}\rho_{nn}(t) + \Jhop{nn,ss'}\rho_{ss'}(t) \label{eq:liouvilletdp1}\\
    &= \Jdyn{nn}\rho_{nn}(t) + \Jhop{nn,ss'}\int_{-\infty}^t{e^{\left(\Jdyn{ss'} - i\tilde{\omega}_{ss'}\right)(t-t')} \Jhop{ss',mm}\rho_{mm}(t')dt'} + O\left({(J^{\text{hop}})}^3\right) \label{eq:liouvilletdp2}
\end{align}
See Appendix \ref{sec:tdpappendix} for a derivation. Note that to simplify our analysis, in Eq.~\eqref{eq:liouvilletdp2}, we have assumed that initially the density matrix is indeed diagonal (i.e., $\rho_{ss'}(-\infty)=0$ for $s\ne s'$). It should be possible to construct a similar analysis including initial off-diagonal terms and terms in higher orders of $J^{\text{hop}}$, however that approach would be much more tedious and we believe the final result will not affect our conclusions.

%In general, Tully-style surface hopping procedures do not explicitly propagate the off-diagonal density matrix elements along their own trajectory, and they do not treat any off-diagonal momentum rescaling either.

The $ss'$ terms in Eq.~\eqref{eq:liouvilletdp2} are summed over all possible states with a nonzero interacting element. Depending on whether or not $m=n$, there are different possible $ss'$ permutations:
\begin{enumerate}
\item If $m\ne n$, there are only two possible choices of $s,s'$ in Eq.~\eqref{eq:liouvilletdp2}: $ss'=nm$ and $ss'=mn$. 
\item If $m=n$,  $ss'$ can equal $ns$ or $sn$ with $s$ representing all possible states that are different from $n$.
\end{enumerate}
To further simplify Eq.~\eqref{eq:liouvilletdp2}, we can use the symmetries of the propagator elements. By analyzing Eqs.~\eqref{eq:Ldyn} and \eqref{eq:Lhop}, we find $\Jdyn{nm}=\Jdyn{mn}$, $\tilde{\omega}_{mn}=-\tilde{\omega}_{nm}$ and $\Jhop{nn,nm}={(\Jhop{nn,mn})}^*=\Jhop{mn,mm}={(\Jhop{nm,mm})}^*$. 
Utilizing these equalities and separating the $m=n$ case from the $m\ne n$ case, Eq.~\eqref{eq:liouvilletdp2} can be evaluated as 
\begin{align}
    \pdv{\rho_{nn}(t)}{t}&\approx \Jdyn{nn}\rho_{nn}(t) + \drhodthop + \drhodtcoh \label{eq:liouvilletdp3}
\end{align}
where (note that there is now only one dummy index $m$ and we have written the summation explicitly)
\begin{align}
    \drhodthop &\equiv \sum_{m\ne n}{2\Re\left[\Jhop{nn,nm}\int_{-\infty}^t{e^{\left(\Jdyn{nm} - i\tilde{\omega}_{nm}\right)(t-t')} {(\Jhop{nn,nm})}^*\rho_{mm}(t')dt'}\right]} \label{eq:drhodthop} \\
    \drhodtcoh &\equiv \sum_{m\ne n}{ 2\Re\left[\Jhop{nn,nm}\int_{-\infty}^t{e^{\left(\Jdyn{nm} - i\tilde{\omega}_{nm}\right)(t-t')} \Jhop{nm,nn}\rho_{nn}(t')dt'}\right]} \label{eq:drhodtcoh}
\end{align}
%\begin{align}
%    \pdv{\rho_{nn}(t)}{t}&\approx \Jdyn{nn}\rho_{nn}(t) + \sum_{m\ne n}{2\Re\left[\Jhop{nn,nm}\int_{-\infty}^t{e^{\left(\Jdyn{nm} - i\omega_{nm}\right)(t-t')} {(\Jhop{nn,nm})}^*\rho_{mm}(t')dt'}\right]} \nonumber\\
%    &\qquad\qquad\qquad + \sum_{m\ne n}{ 2\Re\left[\Jhop{nn,nm}\int_{-\infty}^t{e^{\left(\Jdyn{nm} - i\omega_{nm}\right)(t-t')} \Jhop{nm,nn}\rho_{nn}(t')dt'}\right]} \label{eq:liouvilletdp3}
%\end{align}
%For notational convenience, we define the second term in Eq.~\eqref{eq:liouvilletdp3} as $\drhodthop$ and the last term as $\drhodtcoh$. Therefore Eq.~\eqref{eq:liouvilletdp3} can be written in a compact form as:

In Eq.~\eqref{eq:liouvilletdp3}, the $\Jdyn{nn}$ term captures the propagation of the electronic density matrix within a strict Born-Oppenheimer framework. The $\drhodthop$ term incorporates all the hopping and rescaling as in surface hopping, and we will show below in Sec.~\ref{sec:hop1} that this term is consistent with a normal surface hopping algorithm with a rescaling direction of $\Re[\db_{mn}(\Pb\cdot\db_{mn}^*)]$. The $\drhodtcoh$ term represents population sent to the off-diagonal ($\rho_{nm}$) and then back to the same surface. Since all surface hopping algorithms inspired by Tully allow hops only between different surfaces (not coherences), there is simply no way to capture the effect on nuclear motion caused by this term. We will show below in Sec.~\ref{sec:hop2} that in the adiabatic limit, this term leads to the Berry force proposed in Ref.~\cite{mead1979determination,berry1993Chaotic,subotnik2019demonstration}.

\subsubsection{Momentum Rescaling} \label{sec:hop1}

Within a standard surface hopping approach, a trajectory hopping from surface $m$ to a different surface $n$ corresponds to an exchange of (diagonal) populations within a QCLE formalism ($\rho_{mm}$ to $\rho_{nn}$).
By defining
\begin{align}
    \Delta P_{mn}\equiv \frac{M(E_m-E_n)\db_{mn}}{\Pb\cdot\db_{mn}} \label{eq:deltaphalf}
\end{align}
and utilizing the Hellman-Feynman theorem, we can then write down a suggestive form for the coupling element: 
\begin{align}
    \Jhop{nn,nm} &= d_{mn}^\alpha\frac{P_\alpha}{M}-\frac{1}{2}F_{mn}^\alpha\pdv{}{P_\alpha} =\db_{mn}\cdot\frac{\Pb}{M}\left(1-\frac{1}{2} \Delta P^\alpha_{mn}\pdv{}{P_\alpha}\right) \label{eq:Lhop2}
\end{align}
By employing the ``momentum jump approximation'' \cite{kapral1999Mixed} $1+\alpha\pdv{}{P}= e^{\alpha\pdv{}{P}}+O(\alpha^2)$, we may write Eq.~\eqref{eq:Lhop2} as
\begin{align}
    \Jhop{nn,nm} &=\left(\db_{mn}\cdot\frac{\Pb}{M}\right)e^{-\frac{1}{2}\Delta P^\alpha_{mn}\pdv{}{P_\alpha}} + O({(\Delta P_{mn})}^2) \label{eq:Lhop3}
\end{align}
Plugging Eq.~\eqref{eq:Lhop3} into Eq.~\eqref{eq:drhodthop}, we find
\begin{align}
\begin{split}
    \drhodthop&=2\Re \left[(\db_{mn}\cdot\frac{\Pb}{M})e^{-\frac{1}{2}\Delta P^\alpha_{mn}\pdv{}{P_\alpha}} \right. \\
    \qquad\qquad &\left.\times\int_{-\infty}^t{e^{\left(\Jdyn{nm} - i\tilde{\omega}_{nm}\right)(t-t')} (\db_{mn}^*\cdot\frac{\Pb}{M})e^{-\frac{1}{2}{(\Delta P^\alpha_{mn})}^*\pdv{}{P_\alpha}}\rho_{mm}(t')dt'}\right] + O({(\Delta P_{mn})}^2)
\end{split}\label{eq:drhodthop2}
\end{align}
To further connect with surface hopping, we utilize the fact that $e^{\alpha \pdv{}{P}}f(P)g(P) = (e^{\alpha \pdv{}{P}}f(P))(e^{\alpha \pdv{}{P}}g(P))$ and the fact that, if $\alpha$ and $\beta$ are functions of $P$, $e^{\alpha\pdv{}{P}}e^{\beta\pdv{}{P}} = e^{(\alpha+\beta)\pdv{}{P}} + O(\alpha\beta)$. By applying the factor $e^{-\frac{1}{2}\Delta P_{mn}^\alpha\pdv{}{P_\alpha}}$ in Eq.~\eqref{eq:drhodthop2} on all operators to its right, we find:
\begin{align} \label{eq:drhodthop3}
\begin{split}
    \drhodthop&= 2\Re \left[(\db_{mn}\cdot\frac{\Pb}{M})\int_{-\infty}^t{e^{\left(\Jdyn{nm}(\Pb-\Delta\Pb_{mn}/2)- i\tilde{\omega}_{nm}\right)(t-t')}} \right. \\
     \qquad &\left. \times \left(\mathbf{d}_{mn}^*\cdot\frac{\Pb-\Delta\Pb_{mn}/2}{M}\right)e^{-\Re[\Delta P^\alpha_{mn}]\pdv{}{P_\alpha}}\rho_{mm}(t')dt'\right] + O((\Delta P_{mn})^2)
\end{split}
\end{align}

Eq.~\eqref{eq:drhodthop3} implies that when a diagonal element $\rho_{mm}$ (a population) transforms into $\rho_{nn}$, the total momentum shift is $\Re[\Delta\Pb_{mn}]$ (where $\Delta\Pb_{mn}$ is defined in Eq.~\eqref{eq:deltaphalf}):
\begin{align}
    \Re[\Delta\Pb_{mn}]=M(E_m-E_n)\frac{\Re[\db_{mn}(\Pb\cdot\db_{mn}^*)]}{\abs{\Pb\cdot{\db_{mn}}}^2} \label{eq:deltap}
\end{align}
Therefore, for a surface hopping simulation, the correct rescaling direction for a hop from surface $m$ to $n$ is $\Re[\db_{mn}(\Pb\cdot\db_{mn}^*)]$. When $\db_{mn}$ is real-valued, this direction reduces to $\mathbf{d}_{mn}$ itself, which is consistent with the standard surface hopping algorithm (and as predicted by Pechukas\cite{pechukas1969Timedependent} and Herman\cite{herman1984Nonadiabatic}). 

Let us now show that this hop is indeed consistent with energy conservation (as one usually assumes in surface hopping) to first order in $\hbar$. Note that unlike FSSH, the total energy in PSSH is a non-quadratic function of $\Pb$, so we can only approximate $E_n(\Rb,\Pb+\Delta\Pb_{mn})$ by Taylor expansion to the first order:
\begin{align}
    E_n(\Rb,\Pb+\Re[\Delta\Pb_{mn}])&\approx E_n(\Rb,\Pb)+\pdv{E_n}{P_\alpha}\Re[\Delta P^\alpha_{mn}] \nonumber\\ 
    &= E_n(\Rb,\Pb)+\frac{P_\alpha-i\hbar D_\alpha}{M}\Re[\Delta P^\alpha_{mn}]
\end{align}
As above, we assume $\mathbf{P}$ is large and drop the $\hbar\mathbf{D}$ term. Therefore,
\begin{align}
    E_n(\Rb,\Pb+\Re[\Delta\Pb_{mn}])&\approx E_n(\Rb,\Pb)+\frac{P_\alpha}{M}\Re[\Delta P^\alpha_{mn}] \nonumber\\
    &=E_n(\Rb,\Pb)+(E_m(\Rb,\Pb)-E_n(\Rb,\Pb))\frac{\Re[(\Pb\cdot\db_{mn}^*)(\Pb\cdot\db_{mn})]}{\abs{\Pb\cdot{\db_{mn}}}^2} \nonumber\\
    &=E_m(\Rb,\Pb) \label{eq:rescale}
\end{align}
where in the second to last step, we have used Eq.~\eqref{eq:deltap}. 

\subsubsection{The Coherence Term and the Berry Force} \label{sec:hop2}

We have just shown that, according to the QCLE, the clear semiclassical interpretation of the $\drhodthop$ term is that one must hop between surfaces and rescale momenta accordingly. Next, let us address the term $\drhodtcoh$ (Eq.~\eqref{eq:drhodtcoh}), which describes a diagonal element that hops to an off-diagonal position and then returns to its original position at a later time. For such a process, there will be a change in momentum when the Hamiltonian is complex-valued, but such a change cannot be captured by any standard Tully-style surface hopping approach. 

Since the term $\drhodtcoh$ has no simple hopping interpretation, in order to understand such a term intuitively, we will need to restrict ourselves to the adiabatic limit where we can make several approximations (and will eventually find that a Berry force emerges). First, in the adiabatic limit, we assume the $\abs{\tilde{\omega}_{nm}} \gg \abs{\Jdyn{nm}}$, so that we can drop the dynamical propagation. Second, we make a Markovian approximation so that the integral is dominated by integration of $t'$ close to $t$ and we can replace $\rho(t')$ by $\rho(t)$.
We can then rewrite Eq.~\eqref{eq:drhodtcoh} as:
\begin{align}
    \drhodtcoh&\approx \sum_{m\ne n}{2\Re\left[\Jhop{nn,nm}\Jhop{nm,nn}\int_{-\infty}^t{ e^{-i\tilde{\omega}_{nm}(t-t')}\rho_{nn}(t)dt'}\right]} \nonumber \\
    &= \sum_{m\ne n}{ 2\Re\left[\frac{1}{i\tilde{\omega}_{nm}}\Jhop{nn,nm}\Jhop{nm,nn}\right]}\rho_{nn}(t) \label{eq:drhodtcoh2}
\end{align}
Third, in the adiabatic limit, we can approximate $\abs{(\db_{nn}-\db_{mm})\cdot\Pb}/M \ll \abs{E_n-E_m}$, and therefore we can replace $\tilde{\omega}_{nm}$ by $(E_n-E_m)/\hbar$:
\begin{align}
    \drhodtcoh &\approx \sum_{m\ne n}{ 2\Re\left[\frac{\hbar}{i(E_n-E_m)}\Jhop{nn,nm}\Jhop{nm,nn}\right]}\rho_{nn}(t) \label{eq:drhodtcoh3}
\end{align}
By substituting in the definitions for $\Jhop{nn,nm}$ and $\Jhop{nm,nn}$ in Eq.~\eqref{eq:Lhop}, we have
\begin{align}
    \drhodtcoh
    &\approx \sum_{m\ne n}{\frac{2\hbar}{E_n-E_m}\Im\left[ (\db_{mn}\cdot\frac{\Pb}{M})(1-\frac{1}{2}\Delta P^\alpha_{mn}\pdv{}{P_\alpha})\right.} \nonumber\\
    &\qquad \times \left.(\db_{nm}\cdot\frac{\Pb}{M})(1+\frac{1}{2}(\Delta P^\alpha_{mn})^*\pdv{}{P_\alpha})\right]\rho_{nn}(t) \nonumber\\
    &= \sum_{s\ne n}{\frac{2\hbar(\db_{mn}\cdot\frac{\Pb}{M})(\db_{nm}\cdot\frac{\Pb}{M})\Im[\Delta P_{mn}^\alpha]}{E_m-E_n}\pdv{\rho_{nn}(t)}{P_\alpha}} + O((\Delta P_{mn})^2)
\end{align}
Finally, by plugging in $\Delta \Pb_{mn}$ from Eq.~\eqref{eq:deltaphalf} and dropping the $\O((\Delta P_{mn})^2)$ terms, we arrive at
\begin{align}
    \drhodtcoh &\approx 2\hbar\sum_{m\ne n}{\Im\left[d_{mn}^\alpha(\frac{\Pb}{M}\cdot\db_{nm})\right]}\pdv{\rho_{nn}(t)}{P_\alpha} \nonumber\\
    &= -2\hbar\sum_{m\ne n}{\Im\left[d_{nm}^\alpha(\frac{\Pb}{M}\cdot\db_{mn})\right]}\pdv{\rho_{nn}(t)}{P_\alpha} \label{eq:berryforce}
\end{align}
Eq.~\eqref{eq:berryforce} is exactly the Lorentz-like Berry force term found in Ref.\cite{berry1993Chaotic,subotnik2019demonstration}. If all of the derivative couplings $\mathbf{d}_{nm}$ were real-valued, the Berry force would be zero. However, for a general complex-valued Hamiltonian, the Berry force does not vanish and leads to an additional momentum shift for motion along a fixed adiabatic surface -- which, again, surface hopping is not able to capture. Note that, outside of the adiabatic limit, the meaning of such Berry force effects is complicated and capturing such affects is an outstanding goal for modern semiclassical theories. Note also that such Berry forces can be found using any flavor of a QCLE; it is not an artifact of our introducing any preconditioning for the P-QCLE.

In conclusion, for a system where $\mathbf{d}$ can be made real-valued, we can find that a consistency between a PD-PSSH scheme and the P-QCLE can be achieved, and the correct rescaling direction for a hop from surface $m$ to $n$ is $\mathbf{d}_{mn}$. Thus, obviously, one of the reasons to work with a P-QCLE (rather than a QCLE) is to transform a complex-valued operator into a real-valued operator in phase space. For models with complex-valued derivative couplings, a hopping direction can be formally chosen as $\Re[\mathbf{d}_{nm}(\Pb\cdot\mathbf{d}_{mn})]$; however the effect of the term $\drhodtcoh$ is missing. In particular, in the adiabatic limit, Berry force effects will not be included. This finding is consistent with our observation in Ref.~\cite{miao2019extension}.
%Therefore, one criteria of the pseudo-diabats is that the $\mathbf{d}$ should be as real as possible.

\subsection{Consistency of The Dynamical Propagation and Decoherence} \label{sec:dyn}

When $n=n'$, the dynamical propagator reads
\begin{align}
    \Jdyn{nn} = -\frac{P_\alpha-i\hbar D^\alpha_{nn}}{M}\pdv{}{R_\alpha}-F_{nn}^\alpha\pdv{}{P_\alpha} \label{eq:Ldyn1}
\end{align}
which is consistent with the surface hopping dynamics in Eqs.~\eqref{eq:shr} and \eqref{eq:shp} with active surface $\lambda=n$. If $n\ne n'$, or $n=n'\ne\lambda$, a surface hopping trajectory estimates the force and velocity by its active surface $\lambda$, i.e., it approximates $J^{\text{dyn}}$ by
\begin{align}
    \Jnn{dyn,SH}=-\frac{P-i\hbar D_{\lambda\lambda}^\alpha}{M}\pdv{}{R_\alpha}-F_{\lambda\lambda}^\alpha\pdv{}{P_\alpha} 
\end{align}
According to Refs.~\cite{subotnik2011new,subotnik2013Can,kapral2016Surface}, the difference between $\Jnn{dyn,SH}$ and the actual QCLE propagator $\Jnn{dyn}$ can lead to decoherence. The decoherence rate of $\rho_{nn'}$ is defined by:
\begin{align}
    \gamma_{nn'} &\equiv \frac{\left(\Jnn{dyn}-\Jnn{dyn,SH}\right)\rho_{nn'}}{\rho_{nn'}} \nonumber\\ &=-\frac{1}{2}(\delta D_{nn}^\alpha+\delta D_{n'n'}^\alpha)\frac{1}{\rho_{nn'}}\pdv{\rho_{nn'}}{R_\alpha}
    -\frac{1}{2}(\delta F_{nn}^\alpha+\delta F_{n'n'}^\alpha)\frac{1}{\rho_{nn'}}\pdv{\rho_{nn'}}{P_\alpha} \label{eq:decoherence}
\end{align}
where $\delta D_{nn}^\alpha=-i\hbar (D_{nn}^\alpha-D_{\lambda\lambda}^\alpha)/M$ and $\delta F_{nn}^\alpha=F_{nn}^\alpha-F_{\lambda\lambda}^\alpha$. 
The second term in the RHS of Eq.~\eqref{eq:decoherence} is a standard expression for decoherence; since the early work of Rossky \cite{schwartz1996Quantum,prezhdo1997Evaluation} and Truhlar \cite{jasper2005Electronic}, it has been well known that a decoherence correction for FSSH must be proportional to the difference in forces between different surfaces as these different forces lead to wavepacket separation. Over the last decade, our research group has sought to estimate this term using a so-called AFSSH algorithm \cite{subotnik2013Can} whereby we aim to estimate the factor that multiplies the $\delta F$ term using the history of a given trajectory.

Compared with previous estimates of decoherence times, according to Eq.~\eqref{eq:decoherence}, there is now one new additional term proportional $\delta D$, which corresponds to an new channel for decoherence as induced by the difference in vector potentials. This term is on the order of $\hbar$, which we expect will usually be much smaller than the $\delta F$ term, and so therefore our expectation is that most existing frameworks for treating decoherence (including AFSSH) should apply in general to the current phase space surface hopping approach as well. That being said, in certain cases, especially when the adiabatic force differences are small, one can imagine scenarios whereby decoherence might be induced by the difference in vector potentials, for example, the wavepacket splitting in singlet-triplet crossings \cite{bian2021Modeling}. The consequence of such decoherence will require further investigation.\cite{decoherence}

\section{Discussion}
\subsection{Why and When Will Pseudo-Diabatic PSSH Be Accurate} \label{sec:pseudodiabat}

\subsubsection{The Pseudo-Diabatic PSSH Algorithm}

In Sec.~\ref{sec:pssh}, we have shown that a surface hopping algorithm can be consistent with the QCLE only if a slew of conditions are satisfied, including the fact that the derivative coupling must real-valued. For standard (adiabatic) fewest-switches surface hopping, such a choice will not be possible for a complex-valued Hamiltonian if the real and imaginary parts of the derivative coupling point in different directions. We have recently sought to address this problem in Refs.~\cite{wu2022phasespace,bian2022Modeling} through the present pseudo-diabatic PSSH (PD-PSSH) approach. To best illustrate how the method works, consider a Hamiltonian of the form (in a basis $\ket{j_0}, \ket{j_1}, ...$):
\begin{align}
    h=\begin{bmatrix}h_0&V_1e^{i\phi_1}&V_2e^{i\phi_2}&\ldots\\ V_1e^{-i\phi_1}&h_1&0&0\\ V_2e^{-i\phi_2}&0&h_2&0\\\vdots&0&0&\ddots\\\end{bmatrix} \label{eq:hmodel}
\end{align}
For such a Hamiltonian, we choose a pseudo-diabatic basis of the form 
\begin{align}
  \ket{j} = \begin{cases} \ket{j_0} & j_0 = 0 \\ \ket{j_0}e^{-i\phi_j} & j_0\ne 0 \end{cases} \label{eq:pd}
\end{align}
In such a pseudo-diabatic basis, the Hamiltonian becomes real-valued:
\begin{align}
    h_{\text{PD}}=\begin{bmatrix}h_0&V_1&V_2&\ldots\\ V_1&h_1&0&0\\ V_2&0&h_2&0\\\vdots&0&0&\ddots\\\end{bmatrix}
\end{align}
and the derivative couplings $\mathbf{d}$ between the corresponding phase-space adiabats becomes real-valued as well. This real-valued nature is essential for the surface hopping algorithm to function properly as illustrated in Sec.~\ref{sec:hop2}. Moreover, in this pseudo-diabatic basis, the $\mathbf{D}_W$ matrix reads
\begin{align}
    \mathbf{D}_W =-i\begin{bmatrix}0&0&0&\ldots\\ 0&\nabla\phi_1&0&0\\ 0&0&\nabla\phi_2&0\\\vdots&0&0&\ddots\\\end{bmatrix}
\end{align}
which indicates that the Hamiltonian will have a non-trivial dependence on momentum (assuming the $\nabla \phi$'s are nonzero). As discussed below, these terms result in something akin to a pseudo magnetic-field for the dynamics.

\subsubsection{Inclusion of Nuclear Berry Curvature Effects}

It has been known for a long time that within the realm of nonadiabatic dynamics, the on-diagonal derivative couplings can be regarded as a vector potential \cite{mead1979determination,mead1980molecular,takatsuka2006NonBornOppenheimer}.
The curl of the on-diagonal derivative coupling is called the Berry curvature \cite{berry1993Chaotic,mead1992geometric}, which is analog to a magnetic field in electrodynamics, and it can lead to the Lorentz-like ``Berry force'' for the nuclear motion \cite{berry1993Chaotic,subotnik2019demonstration}.
However, standard surface hopping cannot capture such geometric magnetic effects, as shown in Sec.~\ref{sec:hop2} and in Ref.~\cite{subotnik2019demonstration}. This failure arises because according to the standard QCLE approach, Berry curvature effects arise from the oscillating phase of off-diagonal elements -- which surface hopping cannot maintain correctly.

In the current P-QCLE formalism Eq.~\eqref{eq:pqclead}, geometric effects come from two sources: The coherence effects (which appear in the standard QCLE, see Eq.~\eqref{eq:berryforce}), and the direct vector potential terms $i\hbar\mathbf{D}$ (which appear in $J^{\text{dyn}}$ in Eq.~\eqref{eq:Ldyn}). To better show the effect of the $i\hbar\mathbf{D}$ terms, here we write the equation of motion on a single surface (from Eq.~\eqref{eq:Ldyn}):
\begin{align}
    \left.\pdv{\rho_{nn}}{t}\right|_{\text{single}} = \Jdyn{nn}\rho_{nn} \equiv &-\frac{P_\alpha-i\hbar D_{nn}^\alpha}{M}\pdv{\rho_{nn}}{R_\alpha} \nonumber\\
    &+ \left(\left(\pdv{h}{R_\alpha}\right)_{nn}-\frac{P_\beta-i\hbar D_{nn}^\beta}{M}\pdv{i\hbar D_{nn}^\beta}{R_\alpha}\right)\pdv{\rho_{nn}}{P_\alpha}
     \label{eq:liouvillesingle}
\end{align}
This equation of motion is equivalent to the equation of motion of a swarm of charged particles in a magnetic field. The forces on the particles are (see Appendix \ref{sec:lorentzappendix} for a derivation):
\begin{align}
    f_\alpha = -{\left(\pdv{h}{R_\alpha}\right)}_{nn} + \frac{P_\beta}{M}\left(\pdv{i\hbar D^\beta_{nn}}{R_\alpha}-\pdv{i\hbar D^\alpha_{nn}}{R_\beta}\right) \label{eq:magnetic}
\end{align}
Therefore, altogether, according to a P-QCLE, there appear to be {\em two} magnetic fields operating at the same time (Eqs.~\eqref{eq:berryforce} and \eqref{eq:magnetic} ). In the adiabatic limit (where population on only one surface $n$ is dominant), the total geometric magnetic field that the nuclei feel is then the sum of the fields from the two sources ($\db$ and $\mathbf{D}$):
\begin{align}
    B^{\alpha\beta}_{nn} &= i\hbar\left(\pdv{d^\beta_{nn}}{R_\alpha}-\pdv{d^\alpha_{nn}}{R_\beta} + \pdv{D^\beta_{nn}}{R_\alpha}-\pdv{D^\alpha_{nn}}{R_\beta}\right) \label{eq:magnetic2}
 \end{align}
In practice, when $\db$ is real-valued, all of the field effects are caused by the $i\hbar\mathbf{D}$ term and will be included in PSSH equations of motion (see Eqs.~\eqref{eq:shr} and \eqref{eq:shp}); this realization explains the success of PD-PSSH as far as treating complex-valued Hamiltonians and why it is so important that one keep $\db$ real-valued (or as real-valued as possible). Again, surface hopping will not be able to capture the magnetic field effects as caused by a complex-valued $\db$ tensor.

Lastly, note that, since the phase-space adiabats can be different from the position-space adiabats, the Berry curvature of the phase-space adiabats can also be different from the Berry curvature of the position-space adiabats. 
%However, different from the standard picture of Berry curvature which is only a function of $\Rb$, the $B$ field here depends on both $\Rb$ and $\Pb$. This means that in the PSSH picture, the trajectories are propagated on motion-dependent phase-space adiabats, and they feel a ``dynamical'' Berry curvature. This concept is related with the notion of ``instantaneous field'' proposed by Takatsuka {\em et al} for mean-field dynamics. \cite{takatsuka2010Nonadiabatic}
For example, the following singlet-triplet crossing Hamiltonian
\begin{align}
    h=A\begin{bmatrix}\cos{\theta}&\sin{\theta}e^{i\phi}&\sin{\theta}&\sin{\theta}e^{-i\phi}\\ \sin{\theta}e^{-i\phi}&-\cos{\theta}&0&0\\\sin{\theta}&0&-\cos{\theta}&0\\\sin{\theta}e^{-i\phi} &0&0&-\cos{\theta}\\ \end{bmatrix} \label{eq:h4state}
\end{align}
has zero on-diagonal Berry curvature on all of its position-space adiabats \cite{bian2021Modeling} (so any naive use of a Berry force \cite{bian2022Incorporating} would be ineffectual). However, the on-diagonal Berry curvature is nonzero for the phase-space adiabats if the pseudo-diabats are chosen according to Eq.~\eqref{eq:pd}. For a set of model systems, we have found that our PD-PSSH can successfully model the momentum alterations and reflections encoded by Hamiltonian~\eqref{eq:h4state} \cite{bian2022Modeling}, which implies that the off-diagonal (non-Abelian) part of the Berry curvature can also be captured by PD-PSSH.

%This is indeed align with our previous observations \cite{bian2021Modeling,bian2022Incorporating}, where the nuclear dynamics do have momentum alternation and reflections in Hamiltonian~\eqref{eq:h4state}, very similar to that in models with on-diagonal Berry curvature.

\subsubsection{The Optimal Choice of Basis}

Despite the success of PD-PSSH for certain model problems, it should be noticed that the PD-PSSH algorithm is highly basis-dependent. If the diabats were to be chosen differently for a given Hamiltonian, we would end up with completely different pseudo-diabats and therefore different effective magnetic fields. For example, as pointed in Ref.~\cite{bian2021Modeling}, Hamiltonian~\eqref{eq:h4state} becomes real-valued after a unitary transformation of diabats:
\begin{align}
    h'=A\begin{bmatrix}\cos{\theta}&\sqrt{2}\sin{\theta}\cos{\phi}&\sin{\theta}&\sqrt{2}\sin{\theta}\sin{\phi}\\ \sqrt{2}\sin{\theta}\cos{\phi}&-\cos{\theta}&0&0\\\sin{\theta}&0&-\cos{\theta}&0\\\sqrt{2}\sin{\theta}\sin{\phi} &0&0&-\cos{\theta}\\ \end{bmatrix} \label{eq:h4state2}
\end{align}
Since Hamiltonian~\eqref{eq:h4state2} is already real-valued, the pseudo-diabats prescribed above will equal the diabats, and therefore the resulting phase-space adiabats and position-space adiabats will become equivalent -- each with zero on-diagonal Berry curvature.

Now, a change of basis does not mean that any underlying physics has changed according to exact quantum mechanics. In fact, if one were to run a standard (basis-independent) QCLE simulation for Hamiltonian~\eqref{eq:h4state2}, one expects the wavepackets will still undergo similar separations and reflections arising from coherence effects between different states. That being said, semiclassical PSSH simulations based on Eqs.~\eqref{eq:h4state} and \eqref{eq:h4state2} will yield completely different results (as shown in Fig.~2 in Ref.~\cite{bian2022Modeling}) after a change of basis for two reasons. First, a suboptimal choice of basis can lead to a loss of accuracy by missing out on part of the Berry curvature. Second, a poor choice of basis may also break the unique trajectory assumption (referenced above in Sec.~\ref{sec:pssh}) that is required for meaningful surface hopping dynamics. Obviously, to recover accurate results, a good choice of basis is essential.

%However, Hamiltonian \eqref{eq:h4state2} has two disadvantages: (1) the interpretation of physics is much difficult, since the field terms are not explicitly given; (2) the corresponding PSSH simulations (here they are equivalent to FSSH simulations) will also be less accurate. 

Looking forward, in practice, one would like to run nonadiabatic dynamics for a real molecular system with an {\em ab initio} Hamiltonian--which need not resemble Eq.~\eqref{eq:h4state}. Thus, the question of how to choose the optimal set of pseudo-diabats is a crucial direction for future research; one sorely requires a set of reasonably smooth pseudo-diabats for which most Berry curvature effects can be revealed with PSSH.
% In fact, when $\mathbf{d}_{nn}\equiv 0$, Eq.~\eqref{eq:bc} is the Berry curvature of the phase-space adiabat, see the appendix for the proof. For a two-state Hamiltonian,
% \begin{align}
%     h=A\begin{bmatrix}-\cos{\theta}&\sin{\theta}e^{i\phi}\\\sin{\theta}e^{-i\phi}&\cos{\theta}\\\end{bmatrix}
% \end{align}
% If we choose basis $\ket{\chi_0}=\ket{0}$ and $\ket{\chi_1}=\ket{1}e^{-i\phi}$, the derivative coupling on the ground state phase-space adiabat is
% \begin{align}
%     \mathbf{D}_{00}=-i\nabla\phi\left(1-\frac{A\cos{\theta}}{\sqrt{4A^2+2Ag\cos{\theta}+g^2}}\right)
% \end{align}
% where $g=-\nabla\phi\cdot\frac{\hbar}{m}\Pb+\frac{\hbar^2}{2m}(\nabla\phi)^2$. When $A\to\infty$, we find $B_{00}=-\frac{1}{2}\sin{\theta}\nabla\theta\wedge\nabla\phi$, which are consistent with the expression in Ref. [ref].

\subsection{A Comparison with Shenvi's Adiabatic PSSH Algorithm} \label{sec:shenvi}

At this point, it is appropriate to put the present work in the context of the seminal work of Shenvi. To our knowledge, the first reasonable PSSH algorithm \cite{shenvi2009Phasespace} was proposed by Shenvi, who suggested propagating nuclear dynamics along Berry's so-called superadiabats (as parameterized by both position $\Rb$ and momentum $\Pb$) \cite{berry1987Quantum}. In particular, Shenvi suggested running dynamics through the following approach. First, the nonadiabatic Hamiltonian is transformed to the adiabatic basis:
\begin{align} \label{eq:hasc}
    H_A(\Rb,\Pb) = \frac{{(\Pb-i\hbar\mathbf{d}_A)}^2}{2M} + h_A(\Rb)
\end{align}
where $\mathbf{d}_A$ is the derivative coupling between adiabats and $h_A$ is the adiabatic electronic Hamiltonian (which should be diagonal). Second, $H_A$ is diagonalized, giving the superadiabatic energy $E_S$ and superadiabatic basis. The derivative couplings between superadiabats $\mathbf{d}_S$ (with respect to $\Rb$) and $\bm{\tau}_S$ (with respect to $\Pb$) are calculated in the same way as of the phase-space adiabats (Eqs.~\eqref{eq:drhodp} and \eqref{eq:drhodr}). Third, just like other surface hopping methods, a swarm of trajectories are propagated according to the following equation of motion:
\begin{align}
    \dot{\sigma}&= -\frac{i}{\hbar}\left[E_A(\Rb,\Pb)-i\hbar\dot{\mathbf{R}}\cdot\mathbf{d}_S-i\hbar\dot{\mathbf{P}}\cdot\bm{\tau}_S,\sigma\right] \label{eq:ashsigma} \\
    \dot{\Rb}&=\mel{\lambda}{\nabla_PE^S_\lambda}{\lambda} \label{eq:ashr} \\
    \dot{\Pb}&=-\mel{\lambda}{\nabla_RE^S_\lambda}{\lambda} \label{eq:ashp}
\end{align}

Given the similarity between Eq.~\eqref{eq:hsc} and \eqref{eq:hasc}, it is clear that the A-PSSH dynamics should mimic P-QCLE dynamics if we choose the (positional) adiabats to be the ``pseudo-diabats'' in the preconditioning step. In other words, with all of the caveats mentioned above for the validity of surface hopping, the present paper has shown that Shenvi's A-PSSH can be rationalized as an approximation to the A-QCLE.

Now, as mentioned in Sec.~\ref{sec:introduction}, the standard assumption in quantum dynamics it that the D-QCLE is often superior than the A-QCLE\cite{ryabinkin2014Analysis}. For instance, only the D-QCLE recovers geometric phase around a conical intersection; only the D-QCLE is exact for the spin boson model.\cite{mackernan2002Surfacehopping} Nevertheless, studies by Gherib {\em et al} \cite{gherib2016inclusion} have shown that Shenvi's A-PSSH can capture some diagonal Born-Oppenheimer corrections (DBOC) and therefore can perform better than FSSH when the system is extremely adiabatic. Thus, one might be curious about the performance of the A-QCLE in the extreme adiabatic limit. After all, the previous benchmarks between D-QCLE and A-QCLE (e.g., from Ryabinkin {\em et al}) were performed near conical intersections or normal avoided crossings, which are usually far from the adiabatic limit. One must wonder: will the D-QCLE still outperform the A-QCLE even in the adiabatic limit? The answer is not clear because so much is hidden in these differential equations; we do not know, for example, whether the D-QCLE can capture any DBOC effects indirectly or not.

\begin{figure}[H]
\begin{center}
    \subfloat{\includegraphics[width=0.45\columnwidth]{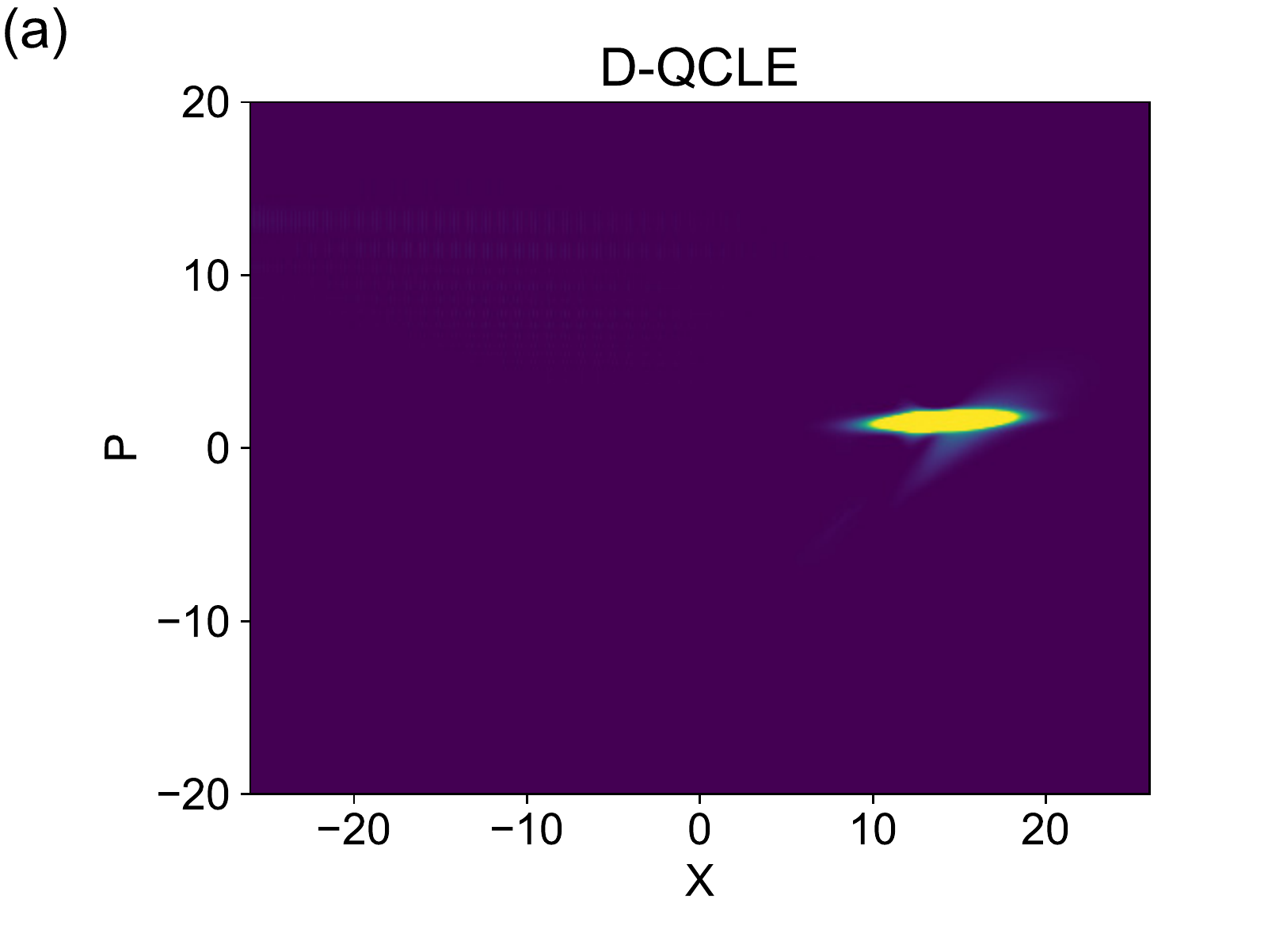}}
    \subfloat{\includegraphics[width=0.45\columnwidth]{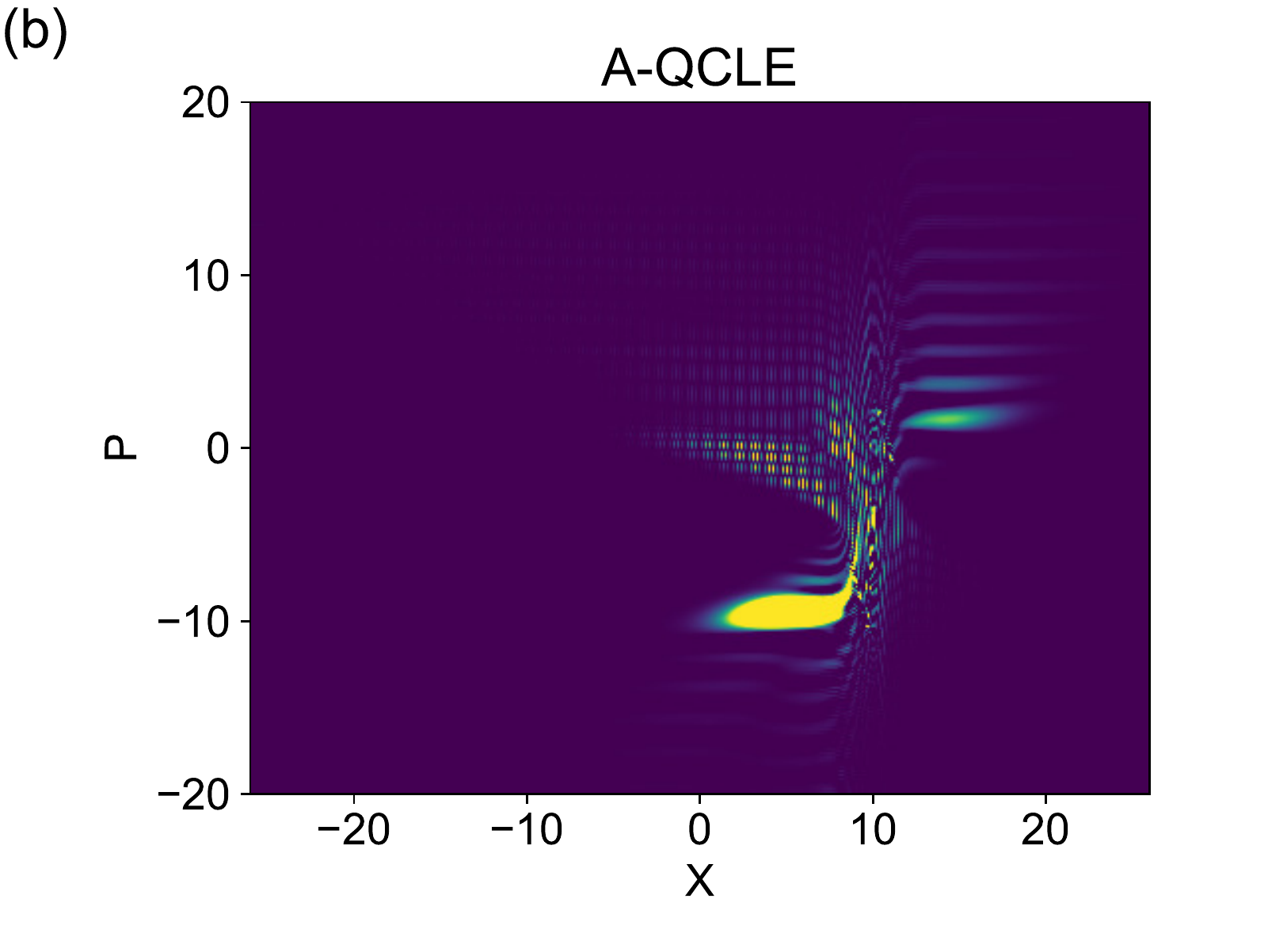}}
    
    \subfloat{\includegraphics[width=0.6\columnwidth]{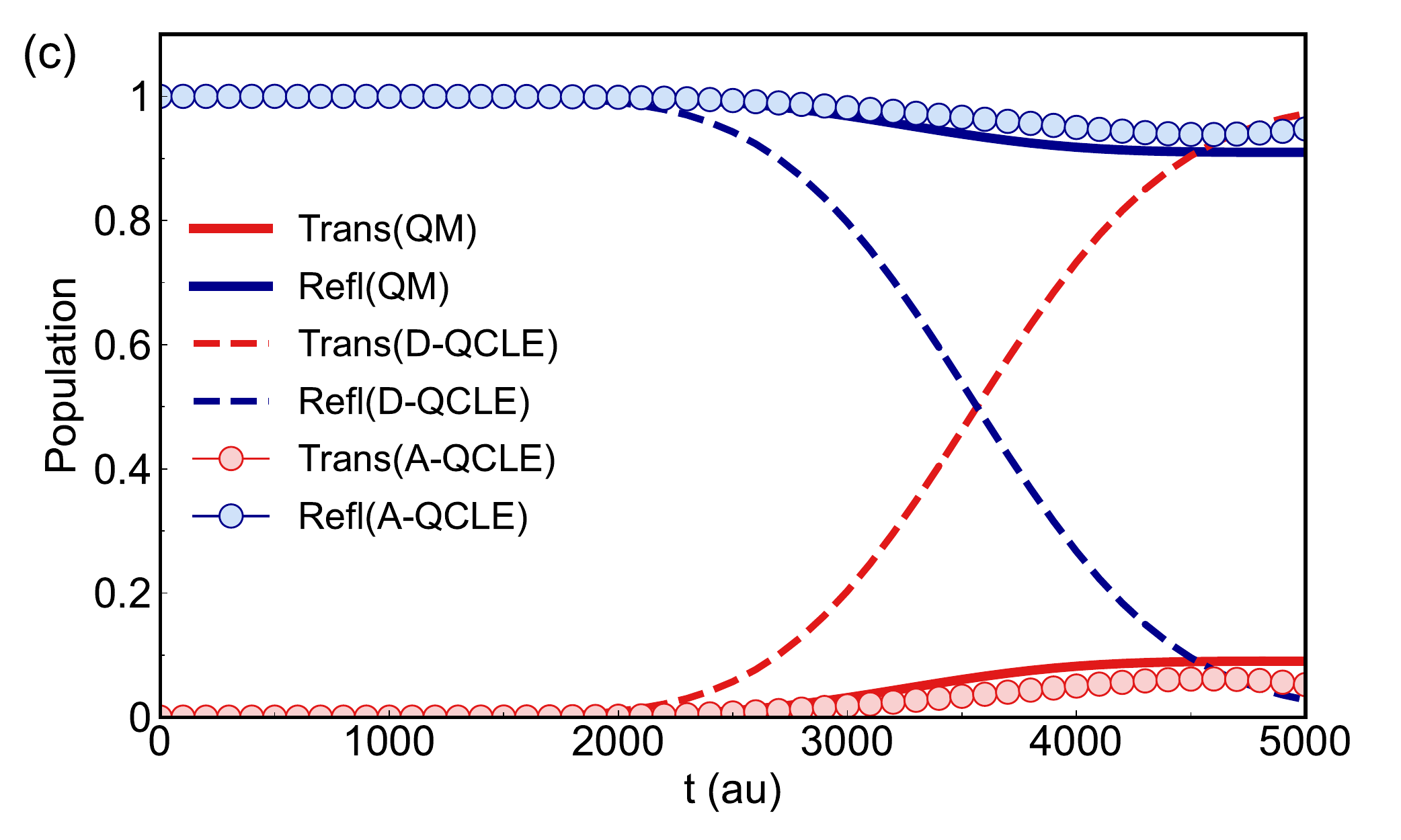}}
    \caption{QCLE and quantum wavepacket simulations of Shenvi's second model, with initial momentum $P_0=5.5$. All parameters are same as the original paper\cite{shenvi2009Phasespace}. (a) (b) Snapshot of the phase-space distribution density matrix $\rho_{11}(\Rb,\Pb)$ in the adiabatic basis at $t=5000 a.u.$, according to (a) the standard QCLE (D-QCLE) and (b) adiabatic-then-Wigner QCLE (A-QCLE). (c) Population as a function of simulation time. The transmitted/reflected populations are calculated by summing over the $X>0$ and $X<0$ regions. Note that both the exact quantum simulation and the A-QCLE predict a high reflection, while D-QCLE predicts almost no reflection. This data confirms that the A-QCLE approach in fact outperforms the D-QCLE in the extreme adiabatic limit (despite the fact that most theory predicts that the D-QCLE should be more accurate\cite{ryabinkin2014Analysis}). There would appear to be no universally best QCLE approach and that the search for an optimal semiclassical theory is not yet complete. More details about the simulation can be found in Appendix~\ref{sec:qcleappendix}.}
    \label{fig:qcle}
    \end{center}
\end{figure}

To find out the answer, we simulated Shenvi's second model (from his original PSSH paper\cite{shenvi2009Phasespace}) on a grid and compared the A-QCLE (Eq.~\eqref{eq:aqcle}) vs the D-QCLE (Eq.~\eqref{eq:dqcle}).
As shown in Fig.~\ref{fig:qcle}, the A-QCLE predicts a significant portion of reflection for initial momentum $P_0 = 5.5$ as the exact result, while the standard D-QCLE has very little reflections. This result indicates that the standard D-QCLE does {\em not} include substantial DBOC effects and therefore is not the optimal choice for this model, at least as compared to the A-QCLE. In other words, the assumption that there is one optimal QCLE is likely not correct; different QCLE's may be optimal for different Hamiltonians and conditions. As a side note, we mention that Shenvi's second model is an extreme model where the diabatic couplings are highly oscillatory; future work will need to assess whether the A-QCLE ever outperforms the standard D-QCLE for a realistic chemical problem of interest.

Interestingly, for a real-valued $2\times 2$ electronic Hamiltonian, Appendix~\ref{sec:shenviappendix} demonstrates that Shenvi's A-PSSH is not only similar to, but equivalent to our PD-PSSH with a certain choice of diabats. Though this equivalence is not general for all Hamiltonians, it suggests that a PD-PSSH are potentially able to capture some DBOC effects (along with geometric magnetic effects) with the proper construction of diabats. In other words, if one chooses an optimal (but as yet unknown) set of diabatic states with which to precondition the Hamiltonian before Wignerization, one can clearly improve the accuracy of a semiclassical simulation.

% For complex-valued systems, Shenvi's PSSH suffers from the gauge problem (since it needs a single-valued adiabat to propagate with), and it is also less obvious how to make $\mathbf{d}$ real-valued, as required by the surface hopping propagation. Nevertheless, it is still interesting to explore the possibility of running Shenvi's PSSH in complex-valued Hamiltonians, since it might be able to capture both the Berry curvature effect and the diagonal Born-Oppenheimer correction.

\subsection{Connection with Mean-Field Ehrenfest Simulations}
Lastly, though the majority of this paper has concentrated on surface hopping, we conclude by emphasizing that the current approach for preconditioning the Hamiltonian also allows one to construct a phase-space Ehrenfest formalism (analogous to surface hopping) with the following equations of motion:
\begin{align}
    \dot{\Pb} &= -\tr_{\text{ele}}[\sigma\nabla_R H_W(\Rb,\Pb)] \label{eq:ef1} \\
    \dot{\Rb} &= \tr_{\text{ele}}[\sigma\nabla_P H_W(\Rb,\Pb)] = \frac{\Pb-i\hbar\tr_{\text{ele}}[\sigma\mathbf{D}_W]}{M} \label{eq:ef2} \\
    \dot{\sigma} &= -\frac{i}{\hbar}[H_W(\Rb,\Pb),\sigma] \label{eq:ef3}
\end{align}
Here, $H_W(\Rb,\Pb) = h_W(\Rb) + \frac{(\Pb-i\hbar\mathbf{D}_W)^2}{2M}$ is the phase-space Hamiltonian defined in Eq.~\eqref{eq:hsc}. The local electronic density matrix $\sigma$ here is defined in the pseudo-diabatic basis. Eqs.~\eqref{eq:ef1}-\eqref{eq:ef3} represent a ``phase-space Ehrenfest'' approach and do not reduce to the usual position-space Ehrenfest theory. 
As an example, we performed Ehrenfest simulations in a complex-valued model system and plot the transmitted and reflected population on different diabats in Fig.~\ref{fig:mfe}; for details see Appendix \ref{sec:mfeappendix}.

\begin{figure}[H]
    \centering
    \includegraphics[width=0.9\textwidth]{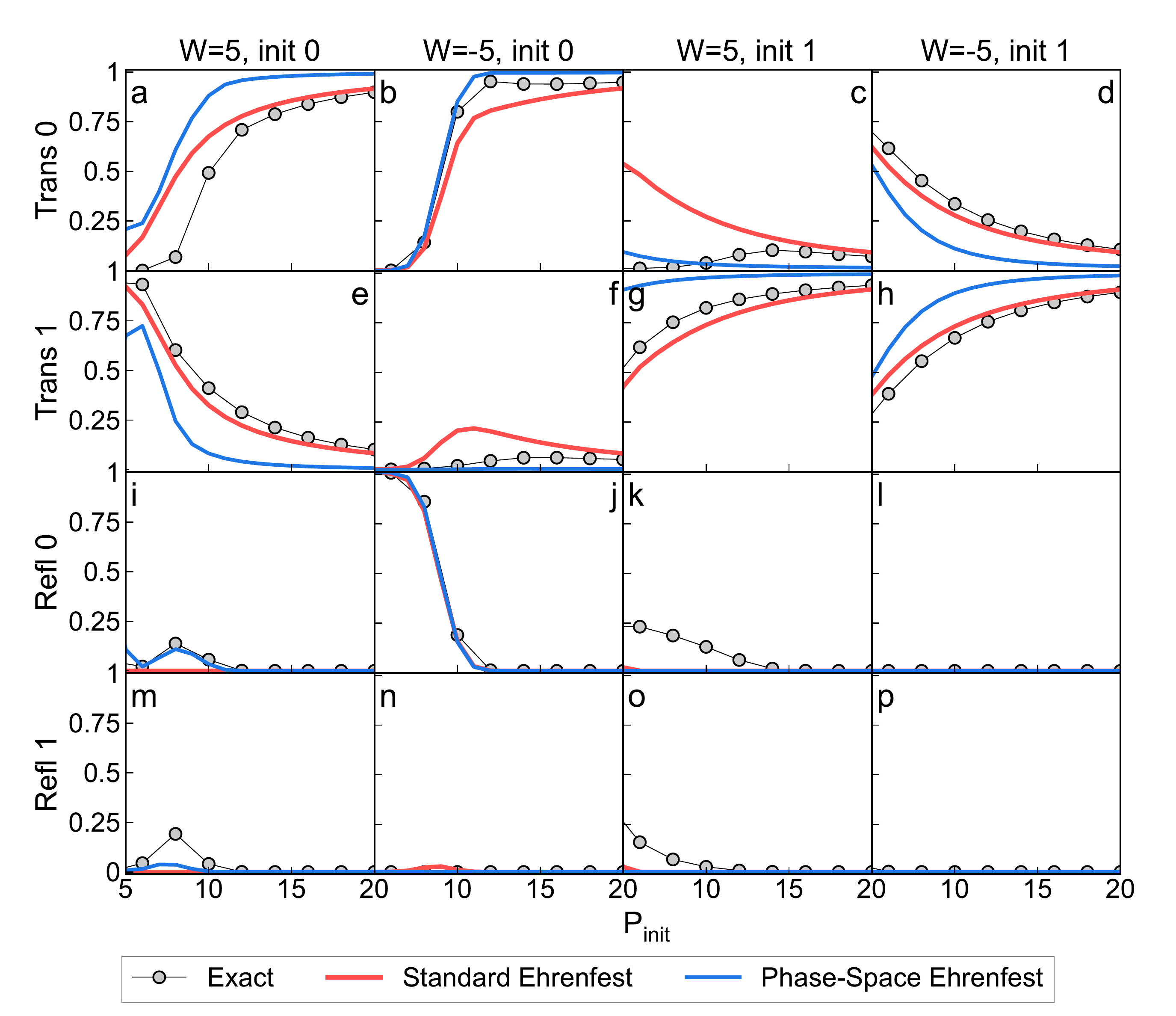}
    \caption{Transmitted and reflected population on different diabats, simulated by  the standard Ehrenfest and phase-space Ehrenfest simulations (Eqs.~\eqref{eq:ef1}-\eqref{eq:ef3}), along with the exact quantum data in the model described in Appendix~\ref{sec:mfeappendix}. We have tested four conditions: $W=5$ and initial diabat $\chi_{\text{init}}=0$ (subfig (a),(e),(i),(m)), $W=-5,\chi_{\text{init}}=0$ (subfig (b),(f),(j),(n)), $W=5,\chi_{\text{init}}=1$ (subfig (c),(g),(k),(o)) and $W=-5,\chi_{\text{init}}=1$ (subfig (d),(h),(l),(p)). For most cases, the conventional Ehrenfest simulation outperforms the phase-space simulation; however, the phase-space Ehrenfest yields a better estimate of the population in subfigures (b) and (i).}
    \label{fig:mfe}
\end{figure}

From the results in Fig.~\ref{fig:mfe}, we find that the conventional Ehrenfest simulation has a better performance on the population in general. However, for certain cases, for example, Fig.~\ref{fig:mfe}b and Fig.~\ref{fig:mfe}i, the phase-space Ehrenfest has a better agreement with the population. As a preliminary test, it is already interesting to see that Ehrenfest has a different behavior by performing the Wignerization on different basis. Interestingly, Cotton and Miller previously proposed an ``adiabatic Ehrenfest'' algorithm where  the classical variables are Wignerized in the adiabatic (rather than diabatic) basis.\cite{cotton2017adiabatic} For nondegenerate real-valued Hamiltonians without Berry curvature (e.g., the two-level spin-boson model, on which they performed the benchmarks), they found the adiabatic Ehrenfest formalism to be equivalent to the standard Ehrenfest.\cite{cotton2017adiabatic}
Clearly, exploring different Ehrenfest approaches for problems of this nature will be very interesting in the future.

\section{Summary}

In this paper, we have shown several results. First, we have demonstrated that a pseudo-diabatic phase-space surface hopping (PD-PSSH) algorithm can be approximately consistent with the quantum-classical Liouville equation (QCLE) picture if one represents the Hamiltonian in a pseudo-diabatic basis and if the derivative couplings between the resulting phase-space adiabats are real-valued. This result justifies our selection of a pseudo-diabatic basis in Ref.~\cite{wu2022phasespace}.
Second and in a similar vein, we have shown that Shenvi's adiabatic PSSH approach \cite{shenvi2009Phasespace} can be mapped to a QCLE that is derived from a Hamiltonian preconditioned by a transformation to the adiabatic basis.
Third, we have demonstrated that, for a complex-valued Hamiltonian, Berry curvature terms (and Berry forces) emerge immediately on the phase-space adiabats according to any QCLE -- however these effect arise from virtual hops between on-diagonal and off-diagonal states and will not be picked up by any FSSH approach.
Fourth, if one preconditions the Hamiltonian through a pseudo-diabatic transformation (such that the final Hamiltonian is real-valued), one can apply a PD-PSSH whereby one propagates the canonical (rather than kinetic) momentum and recover some proper Berry magnetic field effects.
Lastly, although not discussed at length, the formal work here does not make show any significant new physics arising with respect to wavepacket separation and decoherence; we predict that standard decoherence approaches (including AFSSH) should do a reasonable job when applied to complex-valued and/or degenerate Hamiltonians.
Looking forward, one take-home point of this article is clearly that there is no one correct QCLE that is optimal in all dynamical regimes (and therefore there can also be no optimal FSSH approach). In fact, sometimes the A-QCLE does outperform the D-QCLE. Vice versa, we have recently shown that for a set of model Hamiltonians where the optimal preconditioning is obvious,\cite{wu2022phasespace,bian2022Modeling} PD-PSSH can vastly outperform any other surface hopping scheme. 
Thus, the step of finding an optimal basis for preconditioning the dynamics would appear to be essential as we seek to make progress on semiclassical nonadiabatic dynamics and eventually model \abinitio Hamiltonians with spin degrees of freedom and in the presence of magnetic fields.

\section*{Acknowledgments}
This material is based on the work supported by the National Science Foundation under Grant No.~CHE-2102402.

\appendix

\section{Derivation of Eq.~\eqref{eq:liouvilletdp2}} \label{sec:tdpappendix}
\renewcommand{\theequation}{A.\arabic{equation}}

Here we show the derivation of the integral-differential equation Eq.~\eqref{eq:liouvilletdp2} through second order time-dependent perturbation of Eq.~\eqref{eq:liouvilleapprox}. We start by defining
\begin{align}
    \tilde{\rho}_{nn'}(t) = e^{-(\Jnn{dyn}-i\tilde{\omega}_{nn'})t}\rho_{nn'}(t)
\end{align}
Then Eq.~\eqref{eq:liouvilleapprox} reads
\begin{align}
    \pdv{\tilde{\rho}_{nn'}(t)}{t} &= -(\Jnn{dyn}-i\tilde{\omega}_{nn'})e^{-(\Jnn{dyn}-i\tilde{\omega}_{nn'})t}\rho_{nn'}(t) + e^{-(\Jnn{dyn}-i\tilde{\omega}_{nn'})t}\pdv{\rho_{nn'}(t)}{t} \nonumber\\ 
    &= e^{-(\Jnn{dyn}-i\tilde{\omega}_{nn'})t}\Jhop{nn',mm'}\rho_{mm'}(t) \label{eq:appqcle2}
\end{align}
Therefore,
\begin{align}
    \tilde{\rho}_{nn'}(t) &= \int_{-\infty}^{t}{e^{-(\Jnn{dyn}-i\tilde{\omega}_{nn'})t'}\Jhop{nn',mm'}\rho_{mm'}(t')dt'} + \tilde{\rho}_{nn'}(-\infty) 
\end{align}
which is equivalent to
\begin{align}
    \rho_{nn'}(t)= \int_{-\infty}^{t}{e^{(\Jnn{dyn}-i\tilde{\omega}_{nn'})(t-t')}\Jhop{nn',mm'}\rho_{mm'}(t')dt'} + e^{-(\Jnn{dyn}-i\tilde{\omega}_{nn'})t}\rho_{nn'}(-\infty) \label{eq:appqcle3}
\end{align}
By plugging Eq.~\eqref{eq:appqcle3} into the expression for $\rho_{mm'}(t)$ in Eq.~\eqref{eq:liouvilleapprox}, and setting $n=n'$ so that $\tilde{\omega}_{nn'}=0$, we find
\begin{equation}\begin{split}
    \pdv{\rho_{nn}(t)}{t} &= \Jdyn{nn}\rho_{nn}(t) + \Jhop{nn,ss'}e^{-(\Jdyn{ss'}-i\tilde{\omega}_{ss'})t}\rho_{ss'}(-\infty) \\
    &\quad + \Jhop{nn,ss'}\int_{-\infty}^{t}{e^{(\Jdyn{ss'}-i\tilde{\omega}_{ss'})(t-t')}\Jhop{ss',mm'}\rho_{mm'}(t')dt'}
\end{split} \label{eq:appqcle4}
\end{equation}
Because of the form of $J^{\text{hop}}$, $ss'$ must be an off-diagonal term (see Eq.~\eqref{eq:Lhop}), and therefore according to our assumptions in Sec.~\ref{sec:hop}, $\rho_{ss'}(-\infty) \approx 0$ and the second term on the RHS of Eq.~\eqref{eq:appqcle4} can be dropped.

Finally, we will argue that when $m\ne m'$, the last term on the RHS of Eq.~\eqref{eq:appqcle4} can be ignored. To see why, notice that if we plug Eq.~\eqref{eq:appqcle3} into $\rho_{mm'}(t')$ for the last term on the RHS of Eq.~\eqref{eq:appqcle4}, two terms arise:
\begin{equation}\begin{split}
    \pdv{\rho_{nn}(t)}{t} &= \Jdyn{nn}\rho_{nn}(t) + \Jhop{nn,ss'}\int_{-\infty}^{t}{e^{(\Jdyn{ss'}-i\tilde{\omega}_{ss'})(t-t')}\Jhop{ss',mm}\rho_{mm}(t')dt'} \\
    & +\Jhop{nn,ss'}\int_{-\infty}^{t}{e^{(\Jdyn{ss'}-i\tilde{\omega}_{ss'})(t-t')}\Jhop{ss',mm'}dt'\int_{-\infty}^{t'}{e^{(\Jdyn{mm'}-i\tilde{\omega}_{mm'})(t'-t'')}\Jhop{mm',uu'}\rho_{uu'}(t'')dt''}} \\
    & +\Jhop{nn,ss'}\int_{-\infty}^{t}{e^{(\Jdyn{ss'}-i\tilde{\omega}_{ss'})(t-t')}\Jhop{ss',mm'}e^{-(\Jdyn{mm'}-i\tilde{\omega}_{mm'})t'}\rho_{mm'}(-\infty)dt'}
\end{split} \label{eq:appqcle5}
\end{equation}
The third term in RHS of Eq.~\eqref{eq:appqcle5} is third order in $J^{\text{hop}}$ and is truncated by our second order perturbation. The last term of Eq.~\eqref{eq:appqcle5} is also dropped since we have assumed no initial off-diagonal population. By dropping these two terms, we recover Eq.~\eqref{eq:liouvilletdp2}.

\section{Derivation of Lorentz Force from the Continuity Equation \eqref{eq:liouvillesingle}} \label{sec:lorentzappendix}
\renewcommand{\theequation}{B.\arabic{equation}}

Here we show that Eq.~\eqref{eq:liouvillesingle} (with propagation along a single surface $n$) gives a Lorentz-like force along with an adiabatic force. For this subsection only, it will be convenient to define the usual vector potential $\mathbf{A}\equiv i\hbar\mathbf{D}_{nn}$. The key step is to define a density distribution as a function of the kinetic momentum (rather than the canonical momentum):
\begin{align}
    \rho^{\text{kin}}_{nn}(\Rb,\Pb) \equiv \rho_{nn}(\Rb,\Pb+\mathbf{A})
\end{align}
If we propagate this density along the single surface $n$ (as in Eq.~\eqref{eq:liouvillesingle}), we find
\begin{align}
    \left.\pdv{\rho_{nn}^{\text{kin}}(\Rb,\Pb)}{t}\right|_{\text{single}} =& \left.\pdv{\rho_{nn}(\Rb,\Pb+\mathbf{A})}{t}\right|_{\text{single}} \nonumber\\
    = &-\frac{P_\alpha}{M}\left(\pdv{\rho_{nn}}{R_\alpha}\right)_{\Rb,\Pb+\mathbf{A}}
    + \left(\left(\pdv{h}{R_\alpha}\right)_{nn}-\frac{P_\beta}{M}\pdv{A_\beta}{R_\alpha}\right)\left(\pdv{\rho_{nn}}{P_\alpha}\right)_{\Rb,\Pb+\mathbf{A}}
     \label{eq:liouvillesinglekin}
\end{align}
Now, for the derivatives of $\rho^{\text{kin}}_{nn}$, we utilize the chain rule:
\begin{align}
    \left(\pdv{\rho^{\text{kin}}_{nn}}{P_\alpha}\right)_{\Rb,\Pb} &= \left(\pdv{\rho_{nn}}{P_\alpha}\right)_{\Rb,\Pb+\mathbf{A}} \\
    \left(\pdv{\rho^{\text{kin}}_{nn}}{R_\alpha}\right)_{\Rb,\Pb} &= \left(\pdv{\rho_{nn}}{R_\alpha}\right)_{\Rb,\Pb+\mathbf{A}} + \left(\pdv{\rho_{nn}}{P_\beta}\right)_{\Rb,\Pb+\mathbf{A}}\pdv{A_\beta}{R_\alpha}
\end{align}
Plug this equalities into Eq.~\eqref{eq:liouvillesinglekin}, we find
\begin{align}
    \left.\pdv{\rho_{nn}^{\text{kin}}(\Rb,\Pb)}{t}\right|_{\text{single}}
    = &-\frac{P_\alpha}{M}\left(\left(\pdv{\rho^{\text{kin}}_{nn}}{R_\alpha}\right)_{\Rb,\Pb}-\left(\pdv{\rho^{\text{kin}}_{nn}}{P_\beta}\right)_{\Rb,\Pb}\pdv{A_\beta}{R_\alpha}\right) \nonumber\\
    &\qquad + \left(\left(\pdv{h}{R_\alpha}\right)_{nn}-\frac{P_\beta}{M}\pdv{A_\beta}{R_\alpha}\right)\left(\pdv{\rho^{\text{kin}}_{nn}}{P_\alpha}\right)_{\Rb,\Pb} \\
    =&-\frac{P_\alpha}{M}\left(\pdv{\rho^{\text{kin}}_{nn}}{R_\alpha}\right)_{\Rb,\Pb} \nonumber\\
    &\qquad + \left(\left(\pdv{h}{R_\alpha}\right)_{nn}-\frac{P_\beta}{M}\pdv{A_\beta}{R_\alpha}+\frac{P_\beta}{M}\pdv{A_\alpha}{R_\beta}\right)\left(\pdv{\rho^{\text{kin}}_{nn}}{P_\alpha}\right)_{\Rb,\Pb}
    \label{eq:liouvillesinglekin2}
\end{align}
Finally, recall that whenever we express the equation of motion for the density matrix in terms of the position and kinetic (not canonical) momentum, the coefficient of $\pdv{\rho}{P}$ term in a continuity equation represents the force. Therefore, according to the last term in Eq.~\eqref{eq:liouvillesinglekin2}, the true force acting on the system is indeed a Lorentz force plus the adiabatic force in Eq.~\eqref{eq:magnetic}.

\section{Equivalence between $(i)$ Shenvi's Adiabatic PSSH and $(ii)$ the Present Pseudo-diabatic PSSH for Any Real-Valued $2\times 2$ Hamiltonian} \label{sec:shenviappendix}
\renewcommand{\theequation}{C.\arabic{equation}}

Here we show that Shenvi's adiabatic PSSH (A-PSSH) \cite{shenvi2009Phasespace} is equivalent to our pseudo-diabatic PSSH (PD-PSSH) \cite{wu2022phasespace} for real-valued $2\times 2$ Hamiltonians. As mentioned in Sec.~\ref{sec:shenvi}, the only difference between A-PSSH and PD-PSSH are the choices of preconditioned basis: For A-PSSH, we choose the adiabats and for PD-PSSH we choose the pseudo-diabats according to Eq.~\eqref{eq:pd}. Here we show that these two different choices give equivalent pseudo-diabatic Hamiltonians (i.e., the form in Eq.~\eqref{eq:hsc}).

Without loss of generality, we can write any quantum-mechanical nonadiabatic Hamiltonian with a real-valued $2\times 2$ electronic part as
\begin{align}
    H(\Qb) = -\frac{\nabla_Q^2}{2M} + h(\Qb) = -\frac{\nabla_Q^2}{2M} + A\begin{bmatrix}-\cos{\theta}&\sin{\theta} \\ \sin{\theta}&\cos{\theta}\end{bmatrix} \label{eq:h2x2qm}
\end{align}
where both $A(\Qb)$ and $\theta(\Qb)$ are functions of nuclear coordinates. Now by applying a basis transform with
\begin{align}
    U = \frac{1}{\sqrt{2}}\begin{bmatrix}1&-1 \\ i&i\end{bmatrix},
\end{align}
Eq.~\eqref{eq:h2x2qm} becomes
\begin{align}
    H'(\Qb) = -\frac{\nabla_Q^2}{2M} + U^\dagger h(\Qb)U = -\frac{\nabla_Q^2}{2M} + \begin{bmatrix} 0 & Ae^{i\theta} \\ Ae^{-i\theta} & 0\end{bmatrix} \label{eq:h2x2transformed}
\end{align}
Eq.~\eqref{eq:h2x2transformed} is exact. Now, according to Eq.~\eqref{eq:pd}, the pseudo-diabatic basis can be written as the following matrix form:
\begin{align}
    U_{PD}(\Rb) = \begin{bmatrix} 1 & 0 \\ 0 & e^{-i\theta} \end{bmatrix}
\end{align}
Therefore, the pseudo-diabatic Hamiltonian (see Eq.~\eqref{eq:hmodel}) reads
\begin{align}
    H_W^'(\Rb,\Pb) &= \frac{{(\Pb-i\hbar U_{PD}^\dagger \nabla U_{PD})}^2}{2M} + U^\dagger_{PD}U^\dagger h_W(\Rb) UU_{PD} \nonumber\\
    &= \frac{1}{2M}{\begin{bmatrix} \Pb &0\\0&\Pb-\hbar\nabla\theta \end{bmatrix}}^2 + \begin{bmatrix} 0 &A \\ A& 0\end{bmatrix}
\end{align}
To further connect with Shenvi's A-PSSH, we perform a change of variable: we redefine our momentum $\Pb \leftarrow \Pb - \hbar\nabla\theta/2$, therefore $H_W^'$ becomes
\begin{align}
    H_W^'(\Rb,\Pb) = \frac{1}{2M}{\begin{bmatrix} \Pb+\hbar\nabla\theta/2 &0\\0&\Pb-\hbar\nabla\theta/2 \end{bmatrix}}^2 + \begin{bmatrix} 0 &A \\ A& 0\end{bmatrix} \label{eq:pssh_example_1}
\end{align}

Now, in Shenvi's A-PSSH, the total Hamiltonian is Wignerized in the adiabatic basis:
\begin{align}
    U_A(\Rb) = \begin{bmatrix} \cos(\theta/2) & \sin(\theta/2) \\ -\sin(\theta/2) & \cos(\theta/2) \end{bmatrix}
\end{align}
which gives the adiabatic energies  $-A,A$ and adiabatic derivative couplings $\mathbf{d}_A=\nabla\theta/2$. 
Plugging these quantities into Eq.~\eqref{eq:hasc}, we have
\begin{align}
    H_A(\Rb,\Pb) = \frac{{(\Pb - i\hbar\mathbf{d}_A)}^2}{2M} + h_A(\Rb) = \frac{1}{2M}{\begin{bmatrix} \Pb &-i\hbar\nabla\theta/2\\ i\hbar\nabla\theta/2&\Pb \end{bmatrix}}^2 + \begin{bmatrix}-A&0 \\ 0&A\end{bmatrix} \label{eq:pssh_example_2}
\end{align}

Clearly, the Hamiltonian $H_A$ in Eq.~\eqref{eq:pssh_example_2} is equivalent to the Hamiltonian $H'_W$ in Eq.~\eqref{eq:pssh_example_1} up to a constant rotation $U=\begin{bmatrix}-1&i \\ 1&i\end{bmatrix}$. In other words, Shenvi's PSSH superadiabatic Hamiltonian can be captured by one choice of the pseudo-diabats in our phase space approach.
More generally, one can think of the present pseudo-diabatic PSSH algorithm as arising from $(i)$ first applying some transformation to condition the Hamiltonian and $(ii)$ second diagonalizing the Hamiltonian and applying the surface hopping algorithm. Whereas for Shenvi, step $(i)$ always involves diagonalization of the {\em electronic} Hamiltonian (so that step $(ii)$ involves rediagonalization of the {\em total} Hamiltonian), our understanding is that very often, the optimal preconditioning applied in step $(i)$ can be different for different Hamiltonians. Thus, the present algorithm seeks to find the best preconditioning matrix possible. Even though we do not yet have a general approach to picking such a basis and transformation, preliminary evidence suggest that some choices can dramatically improve (or deplete) the accuracy of the final algorithm. 

\section{QCLE Simulations of Shenvi's Model} \label{sec:qcleappendix}
\renewcommand{\theequation}{D.\arabic{equation}}

To compare two flavors of QCLE (A-QCLE and D-QCLE), we use the second model in Shenvi's original paper\cite{shenvi2009Phasespace}. The electronic Hamiltonian is
\begin{align}
    h(X) = A \begin{bmatrix} -\cos{\theta} &\sin{\theta}\\\sin{\theta}&\cos{\theta} \end{bmatrix}
\end{align}
where $X$ is the nuclear degree of freedom, $\theta=\pi C(\tanh(DX)+1)$, $A=0.005$, $C=5.5$ and $D=0.8$ (all in atomic units). The nuclear mass is 2000 $a.u.$. The simulation is performed on a grid basis where $X\in [-36,16]$ is divided into 512 grids and $P\in [-16,24]$ is divided into 384 grids. The semiclassical wavepacket is initialized on the adiabatic one at $X_0=-10$ and $P_0 = 5.5$, with the distribution
\begin{align}
    \rho_0(X,P) = \exp(-2\frac{{(X-X_0)}^2}{\sigma^2} - \frac{\sigma^2{(P-P_0)}^2}{2})
\end{align}
so that it maps to a quantum Gaussian wavepacket with standard deviation $\sigma$. The initial $\sigma$ is chosen to be $20/P_0$. For integration we used a RK2 integrator with a timestep of 0.2 $a.u.$ and a 5-stencil for gradient calculations, so that both the energy and population fluctuation is within 0.1\% throughout our simulation.

Both A-QCLE and D-QCLE simulations are performed on the position-space adiabatic basis. For the A-QCLE, Eq.~(11) in the main text is used. For D-QCLE, Eq.~(10) in the main text is used. For quantum results, we performed a wavepacket simulation using split operator in a 1024 grid with $X\in [-32,32]$.

\section{Ehrenfest Simulations of Our Model} \label{sec:mfeappendix}
\renewcommand{\theequation}{E.\arabic{equation}}

To compare two kinds (conventional and phase-space) of mean-field Ehrenfest simulations, we performed the test on the following model (in a diabatic basis $\ket{0}$ and $\ket{1}$):
\begin{align}
    h(X,Y) = A \begin{bmatrix} -\cos{\theta} &\sin{\theta}e^{i\phi}\\\sin{\theta}e^{-i\phi}&\cos{\theta} \end{bmatrix}
\end{align}
where $X,Y$ are the nuclear degrees of freedom, $\theta=\frac{\pi}{2}(\erf(BX)+1)$, $\phi=WY$, $A=0.02$, $B=3$ and $W=5$ (all in atomic units). The nuclear mass is 1000 $a.u.$. The initial simulation condition corresponds to the Wigner sampling of a quantum wavepacket with expression $\psi_0(X,Y) = \exp(-\frac{{(X-X_0)}^2+{(Y-Y_0)}^2}{\sigma^2}+iP_x(X-X_0)+iP_y(Y-Y_0))$ where $X_0=Y_0=-3$, $P_x = P_y$ and $\sigma=1$. The initial surface is either diabat 0 or diabat 1. The quantum data is reused from the Supplementary Info of Ref.\cite{wu2021Semiclassical}. For both Ehrenfest simulations, we run 10000 trajectories with an integration timestep of 0.05 $a.u.$. The state-wise transmitted and reflected populations (shown in Fig.~\ref{fig:mfe}) are calculated by summing over $\sigma_{nn}$ for each trajectory with final position $X>0$ and $X<0$, respectively.


\begin{thebibliography}{44}
\bibitem[{wig()}]{wigner}
\bibinfo{note}{One might be curious about the whether the Wignerization of the
  operator product $(\Pb\cdot\mathbf{D})_W$ yields any higher order terms. The
  result is negative: As can be seen from the definition of the Moyal bracket
  $e^{-i\hbar\Lambda/2}$, this Wignerization gives no terms above second order
  in $\hbar$, and the first order terms
  $\Pb\hat{\Lambda}\mathbf{D}_W=\sum_\alpha{\nabla_\alpha\mathbf{D}_W}$ and
  $\mathbf{D}_W\hat{\Lambda}\Pb=-\sum_\alpha{\nabla_\alpha\mathbf{D}_W}$ add up
  to zero.}
\bibitem[{com()}]{commutator}
\bibinfo{note}{Note that, if we keep the ${[D_\alpha,[d_\alpha,\rho]]}_+$ term,
  we find a slightly better estimate of the rescaling direction to be
  $\Re[\db_{mn}((\Pb-i\hbar{[\rho,\mathbf{D}]}_+/2)\cdot\db_{mn}^*)]$ and the
  expression for the Berry force becomes $2\hbar\sum_{m\ne
  n}{\Im[\db_{nm}((\Pb-i\hbar\mathbf{D}_{nn})\cdot\db_{mn})/M]}$. Since these
  two values differ from our results in the main text only by a $\hbar$ term
  (see Eqs.~\eqref{eq:deltap} and \eqref{eq:berryforce}), for simplicity, we
  did not use this result in the paper. Furthermore, the
  $\frac{D_\alpha}{M}\pdv{}{R_\alpha}$ term in Eq.~\eqref{eq:Loff} should lead
  to a shift in position (i.e., a ``position rescaling'') of magnitude
  $\hbar\mathbf{D}_{mn}/(\db\cdot\Pb)$ when a hop occurs; however this term
  should have little impact on the dynamics in both the diabatic limit (the
  momentum is large) or the adiabatic limit (the surface is relatively flat),
  and so we have not discussed such a term in detail here either.}
\end{thebibliography}
\end{document}